\documentclass[10pt, conference, compsocconf]{IEEEtran}

\makeatletter
\def\ps@headings{%
\def\@oddhead{\mbox{}\scriptsize\rightmark \hfil \thepage}%
\def\@evenhead{\scriptsize\thepage \hfil \leftmark\mbox{}}%
\def\@oddfoot{}
\def\@evenfoot{}}
\def\blfootnote{\xdef\@thefnmark{}\@footnotetext}
\makeatother

\usepackage{cite}
\usepackage{url}

\usepackage{graphicx}
\usepackage{algorithm}
\usepackage{algorithmic}
\usepackage{subfig}
\usepackage{amssymb, amsmath,graphicx,charter, latexsym}
\usepackage{enumerate}

\newtheorem{definition}{Definition}
\newtheorem{lemma}{Lemma}

\newtheorem{theorem}{Theorem}
\newtheorem{example}{Example}

\def\baselinestretch{0.96}
\begin{document}
\title{Scheduling Periodic Real-Time Tasks with Heterogeneous Reward Requirements}
\author{
\IEEEauthorblockN{I-Hong Hou} \IEEEauthorblockA{CSL and Department
of CS\\University of Illinois\\Urbana, IL 61801,
USA\\ihou2@illinois.edu} \and \IEEEauthorblockN{P. R.
Kumar}\IEEEauthorblockA{CSL and Department of ECE\\University of
Illinois\\Urbana, IL 61801, USA\\prkumar@illinois.edu}
 }
\maketitle\blfootnote{This material is based upon work partially
supported by USARO under Contract Nos. W911NF-08-1-0238 and
W-911-NF-0710287, AFOSR under Contract FA9550-09-0121, and NSF under
Contract No. CNS-07-21992. Any opinions, findings, and conclusions
or recommendations expressed in this publication are those of the
authors and do not necessarily reflect the views of the above
agencies. }

\begin{abstract}
We study the problem of scheduling periodic real-time tasks so as to
meet their individual minimum reward requirements. A task generates
jobs that can be given arbitrary service times before their
deadlines. A task then obtains rewards based on the service times
received by its jobs. We show that this model is compatible to the
imprecise computation models and the increasing reward with
increasing service models. In contrast to previous work on these
models, which mainly focus on maximize the total reward in the
system, we aim to fulfill different reward requirements by different
tasks, which offers better fairness and allows fine-grained tradeoff
between tasks. We first derive a necessary and sufficient condition
for a system, along with reward requirements of tasks, to be
feasible. We also obtain an off-line feasibility optimal scheduling
policy. We then studies a sufficient condition for a policy to be
feasibility optimal or achieves some approximation bound. This
condition can serve as a guideline for designing on-line scheduling
policy and we obtains a greedy policy based on it. We prove that the
on-line policy is feasibility optimal when all tasks have the same
periods and also obtain an approximation bound for the policy under
general cases.
\end{abstract}

\section{Introduction}
\label{section:introduction}

In classical hard real-time systems, every job needs to be completed
before its deadline, or the system suffers from a timing fault. In
practice, many applications allow approximate results and partially
completed jobs only degrade the overall performance rather than
causing a fault. Imprecise computation models \cite{JC90,JL91} and
increasing reward with increasing service (IRIS) models \cite{JD96}
have been proposed to deal with such applications. Most work on
these models only aims to minimize the total error, or,
equivalently, maximize the total reward of the system without any
considerations on fairness. However, in many applications, rewards
of different tasks are not additive and satisfying individual reward
requirements is more important than maximizing total rewards. For
example, consider a server that provides video streams to
subscribers. Deadline misses will only cause losses on some frames
and degrade the video quality, which is usually tolerable as long as
such losses happen infrequently. In such an application, a policy
that aims to maximize total reward may end up providing perfect
video quality for some subscribers while only offering poor quality
for others. In contrast, a desirable policy should aim at providing
reasonably good quality to all of its subscribers.

In this paper, we describe a model that considers the hard delay
bounds of tasks as well as rewards for partially completed jobs, in
a system with a set of periodic tasks. The relationship between
service times and rewards can be any arbitrary increasing and
concave function and may differ from task to task. We allow each
task to have its own individual requirement on the average reward it
obtains. We show that both the imprecise computation model and the
IRIS model are special cases of our model.

Based on the model, we first analyze the conditions for feasibility,
that is, whether there exists a scheduling policy that meets the
individual reward requirements of all tasks in the system. We prove
a necessary and sufficient condition for feasibility. We also
propose a linear time algorithm for evaluating whether a system is
feasible. Along with the feasibility condition, we also derive an
off-line scheduling policy that is feasibility optimal, meaning that
it fulfills all feasible systems.

We then study the problem of designing on-line scheduling policies.
We derive a sufficient condition for a policy to be feasibility
optimal, or, serve as an approximation policy with some
approximation bound. Using this condition as a guideline, we propose
an on-line scheduling policy. We prove that this on-line policy
fulfills every feasible system in which periods are the same for all
tasks. We also obtain an approximation bound for this policy when
periods of tasks may be different.

In addition to theoretical studies, we also conduct simulations to
verify our results. We compare our policy against one proposed by
Aydin et al \cite{HA01}, which is proved to be an optimal off-line
policy that maximizes the total reward in any system. Simulation
results suggest that although the policy proposed by \cite{HA01}
achieves maximum total reward, it can result in severe unfairness
and does not allow fine-grained tradeoffs between the performances
of different tasks.

The rest of the paper is organized as follows. Section
\ref{section:related} summaries some existing work and, in
particular, introduces the basic concepts in the imprecise
computation model and the IRIS model. Section \ref{section:model}
formally describes our proposed model and discusses how it can
capture the imprecise computation model and the IRIS model. Section
\ref{section:feasibility} analyzes the necessary and sufficient
condition for a system to be feasible, and proposes a linear time
algorithm for evaluating feasibility. Section
\ref{section:scheduling} studies the problem of scheduling jobs and
obtains a sufficient condition for a policy to achieve an
approximation bound or to be feasibility optimal. Based on this
condition, Section \ref{section:greedy} proposes a simple on-line
scheduling policy and analyzes its performance under different
cases. Section \ref{section:simulation} demonstrates our simulation
results. Finally, Section \ref{section:conclusion} concludes this
paper.

\section{Related Work}
\label{section:related}

The imprecise computation models \cite{JC90,JL91} have been proposed
to handle applications in which partially completed jobs are useful.
In this model, all jobs consist of two parts: a mandatory part and
an optional part. The mandatory part needs to be completed before
its deadline, or else the system suffers from a timing fault. On the
other hand, the optional part is used to further enhance performance
by either reducing errors or increasing rewards. The relations
between the errors, or rewards, and the time spent on the optional
parts, are described through error functions or reward functions.
Chung, Liu, and Lin \cite{JC90} have proposed scheduling policies
that aim to minimize the total average error in the system for this
model. Their result is optimal only when the error functions are
linear and tasks generate jobs with the same period. Shih and Liu
\cite{WKS95} have proposed policies that minimize the maximum error
among all tasks in the system when error functions are linear.
Feiler and Walker \cite{PF01} have used feedback to
opportunistically schedule optional parts when the lengths of
mandatory parts may be time-varying. Mejia-Alvarez, Melhem, and
Mosse \cite{PMA00} have studied the problem of maximizing total
rewards in the system when job generations are dynamic. Chen et al
\cite{JMC09} have proposed scheduling policies that defer optional
parts so as to provide more timely response for mandatory parts. Zu
and Chang \cite{MZ03} have studied the scheduling problem when
optional parts are hierarchical. Aydin et al \cite{HA01} have
proposed an off-line scheduling policy that maximizes total rewards
when the reward functions are increasing and concave. Most of these
works only concern the maximization of the total reward in a system.
Amirijoo, Hansson, and Son \cite{MA06} have considered the tradeoff
between data errors and transaction errors in a real-time database.
The IRIS models can be thought of as special cases of the imprecise
computation models where the lengths of mandatory parts are zero.
Scheduling policies aimed at maximizing total rewards have been
studied for such models \cite{JD96,HC00}.

\section{System Model}
\label{section:model}

Consider a system with a set $S=\{A,B,\dots\}$ of real-time tasks.
Time is slotted and expressed by $t\in\{0,1,2,\dots\}$. Each task
$X$ generates a job periodically with period $\tau_X$. A job can be
executed multiple times in the period that it is generated; the
execution of a job does not mean its completion. The job is removed
from the system when the next period begins. In other words, the
relative deadline of a job generated by task $X$ is also $\tau_X$.
We assume that all tasks in $S$ generate a job at time $t=0$. We
denote a frame as the time between two consecutive time slots where
all tasks generate a job. The length of a frame, which we denote by
$T$, is the least common multiple of $\{\tau_X|X\in S\}$. Thus, a
frame consists of $T/\tau_X$ periods of task $X$.

As noted above, each job can be executed an arbitrary number of time
slots before its deadline. Each task obtains a certain \emph{reward}
each time that its job is executed. The total amount of reward
obtained by a task in a period depends on the number of times that
its job has been executed in the period. More formally, task $X$
obtains reward $r^{i}_X\geq0$ when it executes its job for the
$i^{th}$ time in a period. For example, if a job of task $X$ is
executed a total of $n$ time slots, then the total reward obtained
by task $X$ in this period is $r^1_X+r^2_X+\dots+r^n_X$. We further
assume that the marginal reward of executing a job decreases as the
number of executions increases, that is, $r^{i+1}_X\leq r^{i}_X$,
for all $i$ and $X$. Thus, the total reward that a task obtains in a
period is an increasing and concave function of the number of time
slots that its job is executed.

\begin{figure}[t]
\label{fig:audio_channel_debt} 
\includegraphics[width=3.2in]{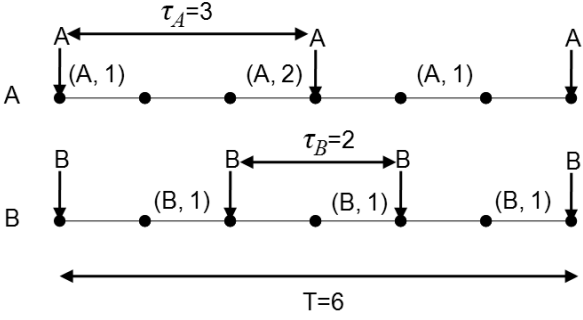}
\caption{An example of the system and scheduling policy over a frame
of six slots, which consists of two period of task $A$ and three
periods of task $B$. Arrows in the figure indicate the beginning of
a new period.} \label{figure:model:example}
\end{figure}

A \emph{scheduling policy} $\eta$ for the system is one that chooses
an \emph{action} in each time slot. The action taken by $\eta$ at
time $t$ is described by $\eta(t)=(X,i)$, meaning that the policy
executes the job of task $X$ at time $t$ and that this is the
$i^{th}$ time that the job is being executed in the period. Fig.
\ref{figure:model:example} shows an example with two tasks over one
frame, which consists of two periods of task $A$ and three periods
of task $B$. In this example, action $(A,1)$ is executed twice and
$(A,2)$ is executed once. Thus, the reward obtained by task $A$ in
this frame is $2r^1_A+r^2_A$. On the other hand, the reward obtained
by task $B$ in this frame is $3r^1_B$.

The performance of the system is described by the long-term
\emph{average reward} per frame of each task in the system.
\begin{definition}
Let $s_X(t)$ be the total reward obtained by task $X$ between time 0
and time $t$ under some scheduling policy $\eta$. The \emph{average
reward} of task $X$ is defined as $q_X:=\liminf_{t\rightarrow
\infty} \frac{s_X(t)}{t/T}$.
\end{definition}

We assume that there is a minimum average reward requirement for
each task $X$, $q^*_X>0$. We wish to verify whether a system is
\emph{feasible}, that is, whether each task can have its minimum
average reward requirement satisfied.
\begin{definition}
A system is \emph{fulfilled} by a scheduling policy $\eta$ if, under
$\eta$, $q_X \geq q^*_X$ with probability 1, for all $X\in S$.
\end{definition}
\begin{definition}
A system is \emph{feasible} if there exists some scheduling policy
that fulfills it.
\end{definition}

A natural metric to evaluate a scheduling policy is the set of
systems that the policy fulfills. For ease of discussion, we only
consider systems that are \emph{strictly feasible}
\begin{definition}
A system with minimum reward requirements $[q^*_X|X\in S]$ is
\emph{strictly feasible} if there exists some $\epsilon>0$ such that
the same system with minimum reward requirements
$[(1+\epsilon)q^*_X]$ is feasible.
\end{definition}
\begin{definition}
A scheduling policy is \emph{feasibility optimal} if it fulfills all
strictly feasible systems.
\end{definition}

Moreover, since the overhead for computing a feasibility optimal
policy may be too high in certain scenarios, we also need to
consider simple approximation policies.
\begin{definition}
A scheduling policy is a $p$-\emph{approximation policy}, $p\geq1$,
if it fulfills all systems with minimum reward requirements
$[q^*_X]$ such that the same system with minimum reward requirements
$[pq^*_X]$ is strictly feasible.
\end{definition}

\subsection{Extensions for Imprecise Computation Models}
\label{section:model:imprecise}

In this section, we discuss how our proposed model can be used to
handle imprecise computation models and  IRIS models. In such
models, a task consists of two parts: a mandatory part and an
optional part. The mandatory part is required to be completed in
each period, or else the system fails. After the mandatory part is
completed, the optional part can be executed to improve performance.
The more optional parts executed for a task, the more rewards it
gets.

Let $m_X$ be the length of the mandatory part of task $X$, that is,
it is required that each job of $X$ is executed at least $m_X$ time
slots in each of its period. Let $o_X$ be the length of the optional
part of task $X$. To accommodate this scenario, we define a symbolic
value $M$ with the following arithmetic reminiscent of the ``Big-$M$
Method'' in linear programming: $0\times M=0$, $aM+bM = (a+b)M$,
$a\times(bM)=(ab)M$, $aM+c>bM+d$, if $a>b$, and $aM+c>aM+d$ if
$c>d$, for all real numbers $a,b,c,d$. Loosely speaking, $M$ can be
thought of as a huge positive number. For each task $X$, we set
$r^{1}_X=r^2_X=\dots=r^{m_X}_X=M$, we then set
$r^{m_X+1}_X,r^{m_X+2}_X,\dots,r^{m_X+o_X}_X$ according to the
rewards obtained by $X$ for its optional part, and $r^{i}_X=0$, for
all $i>m_X+o_X$. The minimum reward requirement of task $X$ is set
to be $\frac{T}{\tau_X}m_XM+\hat{q}^*_X$ with $\hat{q}^*_X\geq 0$.
Thus, a scheduling policy that fulfills such a system is guaranteed
to complete each mandatory part with probability one.

\section{Feasibility Analysis}
\label{section:feasibility}

In this section, we establish a necessary and sufficient condition
for a system to be feasible. Consider a feasible system that is
fulfilled by a policy $\eta$. Suppose that, on average, there are
$f^{i}_X$ periods of task $X$ in a frame in which the action $(X,i)$
is taken by $\eta$. The average reward of task $X$ can then be
expressed as $q_X=\sum_{i=1}^{\tau_X} f^{i}_Xr^{i}_X$. We can
immediately obtain a necessary condition for feasibility.

\begin{lemma}   \label{lemma:feasibility:necessary}
A system with a set of tasks $S=\{A,B,\dots\}$ is feasible only if
there exists $\{f^i_X|X\in S, 1\leq i\leq \tau_X\}$ such that
\begin{align}
&q^*_X \leq \sum_{i=1}^{\tau_X} f^{i}_Xr^{i}_X, &\forall X\in S,\label{equation:feasibility:necessary1}\\
&0\leq f^{i}_X\leq \frac{T}{\tau_X}, &\forall X\in S, 1\leq i\leq
\tau_X,\label{equation:feasibility:necessary2}\\
&\sum_{X\in S}\sum_{i=1}^{\tau_X}f^{i}_X\leq
T.\label{equation:feasibility:necessary3}
\end{align}
\end{lemma}
\begin{IEEEproof}
Condition (\ref{equation:feasibility:necessary1}) holds because task
$X$ requires that $q^*_X\leq q_X$. Condition
(\ref{equation:feasibility:necessary2}) holds because there are
$T/\tau_X$ periods of task $X$ in a frame and thus $f^{i}_X$ is
upper-bounded by $T/\tau_X$. Finally, the total average number of
time slots that the system executes one of the jobs in a frame can
be expressed as $\sum_{X\in S}\sum_{i=1}^{\tau_X}f^{i}_X$, which is
upper-bounded by the number of time slots in a frame, $T$. Thus,
condition (\ref{equation:feasibility:necessary3}) follows.
\end{IEEEproof}

Next, we show that the conditions
(\ref{equation:feasibility:necessary1})--(\ref{equation:feasibility:necessary3})
are also sufficient for feasibility. To prove this, we first show
that the polytope, which contains all points $f=(f^1_A, f^2_A,\dots,
f^{\tau_A}_A, f^1_B,\dots)$ that satisfy conditions
(\ref{equation:feasibility:necessary2}) and
(\ref{equation:feasibility:necessary3}), is a convex hull of several
integer points. We then show that for all integer points $n =
(n^1_A, n^2_A,\dots, n^{\tau_A}_A, n^1_B,\dots)$ in the polytope,
there is a schedule under which the reward obtained by task $X$ is
at least $\sum_{i=1}^{\tau_X} n^i_Xr^i_X$. We then prove sufficiency
using these two results.

Define a matrix $H=[h_{i,j}]$ with $2\sum_{X\in S} \tau_X + 1$ rows
and $\sum_{X\in S}\tau_X$ columns as follows:
\begin{equation}
h_{i,j}=\left \{ \begin{array}{rl}
1, & \mbox{if $i=1$,}\\
1, & \mbox{if $i = 2j$,}\\
-1, & \mbox{if $i = 2j+1$,}\\
0, & \mbox{else.}\end{array}\right.
\end{equation}
Define $b=[b_i]$ to be a column vector with $2\sum_{X\in S} \tau_X +
1$ elements so that $b_1 = T$; the first $\tau_A$ elements with even
indices are set to $T/\tau_A$, that is, $b_2 =
b_4=\dots=b_{2\tau_A}=T/\tau_A$; the next $\tau_B$ elements with
even indices are set to $T/\tau_B$, and so on. All other elements
are set to 0. For example, the system shown in Fig.
\ref{figure:model:example} would have
\[
H=\left [ \begin{array}{rrrrr}
1, &1,&1,&1,&\dots\\
1, &0,&0,&0,&\dots\\
-1,&0,&0,&0,&\dots\\
0, &1,&0,&0,&\dots\\
0,&-1,&0,&0,&\dots\\
0,&0,&1,&0,&\dots\\
0,&0,&-1,&0,&\dots\\ &&\vdots\end{array}\right], b=\left [
\begin{array}{c}
T\\
T/\tau_A\\
0\\
T/\tau_A\\
0\\
\vdots\\
T/\tau_B\\
0\\
\vdots\end{array}\right].
\]
Thus, conditions (\ref{equation:feasibility:necessary2}) and
(\ref{equation:feasibility:necessary3}) can be described as $Hf\leq
b$. Theorem 5.20 and Theorem 5.24 in \cite{BK} shows the following:
\begin{theorem} \label{theorem:feasibility:TUM}
The polytope defined by $\{f|Hf\leq b\}$, where $b$ is an integer
vector, is a convex hull of several integer points if for every
subset $R$ of rows in $H$, there exists a partition of $R=R_1\cup
R_2$ such that for every column $j$ in $H$, we have $\sum_{i_1\in
R_1} h_{i_1,j} - \sum_{i_2\in R_2} h_{i_2, j}\in
\{1,0,-1\}$.\footnote{Such matrix $H$ is called a \emph{totally
unimodular matrix} in combinatorial optimization theory.}
\end{theorem}

Since all elements in $b$ are integers, we obtain the following:
\begin{theorem} \label{theorem:feasibility:integer hull}
The polytope defined by $\{f|Hf\leq b\}$, where $H$ and $b$ are
derived using conditions (\ref{equation:feasibility:necessary2}) and
(\ref{equation:feasibility:necessary3}) as above, is a convex hull
of several integer points.
\end{theorem}
\begin{IEEEproof}
Let $R=\{r_1,r_2,\dots\}$ be the indices of some subset of rows in
$H$. If the first row is in $R$, we choose $R_1 = \{r|r\in R,
\mbox{$r$ is odd}\}$ and $R_2 = \{r|r\in R, \mbox{$r$ is even}\}$.
Since for all columns $j$ in $H$, $h_{1,j}=1$, $h_{2j,j}=1$,
$h_{2j+1,j}=-1$, and all other elements in column $j$ are zero, we
have $\sum_{i_1\in R_1} h_{i_1,j} - \sum_{i_2\in R_2} h_{i_2, j}\in
\{1,0,-1\}$. On the other hand, if the first row is not in $R$, we
choose $R_1=R$ and $R_2=\varnothing$. Again, we have $\sum_{i_1\in
R_1} h_{i_1,j} - \sum_{i_2\in R_2} h_{i_2, j}=\sum_{i_1\in R_1}
h_{i_1,j}\in \{1,0,-1\}$. Thus, by Theorem
\ref{theorem:feasibility:TUM}, the polytope defined by $\{f|Hf\leq
b\}$ is a convex hull of several integer points.
\end{IEEEproof}

Next we show that all integer points in the polytope can be carried
out by some scheduling policy as follows:

\begin{theorem} \label{theorem:feasibility:integer feasible}
Let $n = (n^1_A, n^2_A,\dots, n^{\tau_A}_A, n^1_B,\dots)$ be an
integer point in the polytope $\{f|Hf\leq b\}$. Then, there exists a
scheduling policy so that $q_X\geq \sum_{i=1}^{\tau_X} n^i_Xr^i_X$.
\end{theorem}
\begin{IEEEproof}
We prove this theorem by constructing a scheduling policy that
achieves the aforementioned requirement. We begin by marking
deadlines of actions. Ideally, we wish to schedule the action
$(X,i)$ $n^i_X$ times in a frame. Without loss of generality, we
assume that the frame starts at time $0$ and ends at time $T$. Since
there is at most one $(X,i)$ action in a period of task $X$, we can
mark the deadlines of these actions as $T, T-\tau_X, \dots,
T-(n^i_X-1)\tau_X$, respectively. The scheduling policy will
schedule the action with the earliest deadline that has neither been
executed in its period (that is, it does not schedule two actions of
the same type in the same period) nor missed its deadline, with ties
broken arbitrarily. Fig. \ref{fig:feasibility:edf0} shows how the
deadlines are marked in an example with $n^1_A=1, n^2_A=2, n^1_B=2,$
and $n^2_B=1$. Fig. \ref{fig:feasibility:edf1} shows the resulting
schedule for this example. Note that, under this policy, there are
time slots where the policy schedules an action $(X,i)$, but it is
instead the $j^{th}$ time that the job of $X$ is executed in the
period. Thus, we may need to renumber these actions as in Fig.
\ref{fig:feasibility:edf2} and define $\bar{n}^i_X$ as the actual
number of times that the action $(X,i)$ is executed in the frame. In
the example of Fig. \ref{fig:feasibility:edf}, we have $\bar{n}^1_A
= 2, \bar{n}^2_A = 1, \bar{n}^1_B = 3$, and $\bar{n}^2_B = 0$. Since
the policy does not schedule two identical actions in a period, as
long as all actions are executed before their respective deadlines,
we have $\sum_{i=1}^k\bar{n}^i_X\geq \sum_{i=1}^kn^i_X$, for all $k$
and $X\in S$. Thus, $q_X = \sum_{i=1}^{\tau_X}\bar{n}^i_Xr^i_X\geq
\sum_{i=1}^{\tau_X} n^i_Xr^i_X$, since $r^1_X\geq r^2_X\geq\dots$,
for all $X\in S$. To avoid confusion, we refer actions by its type
before renumbering throughout the rest of the proof.

\begin{figure}[t]
\subfloat[The deadlines that are marked]{
\label{fig:feasibility:edf0} 
\includegraphics[width=3.2in]{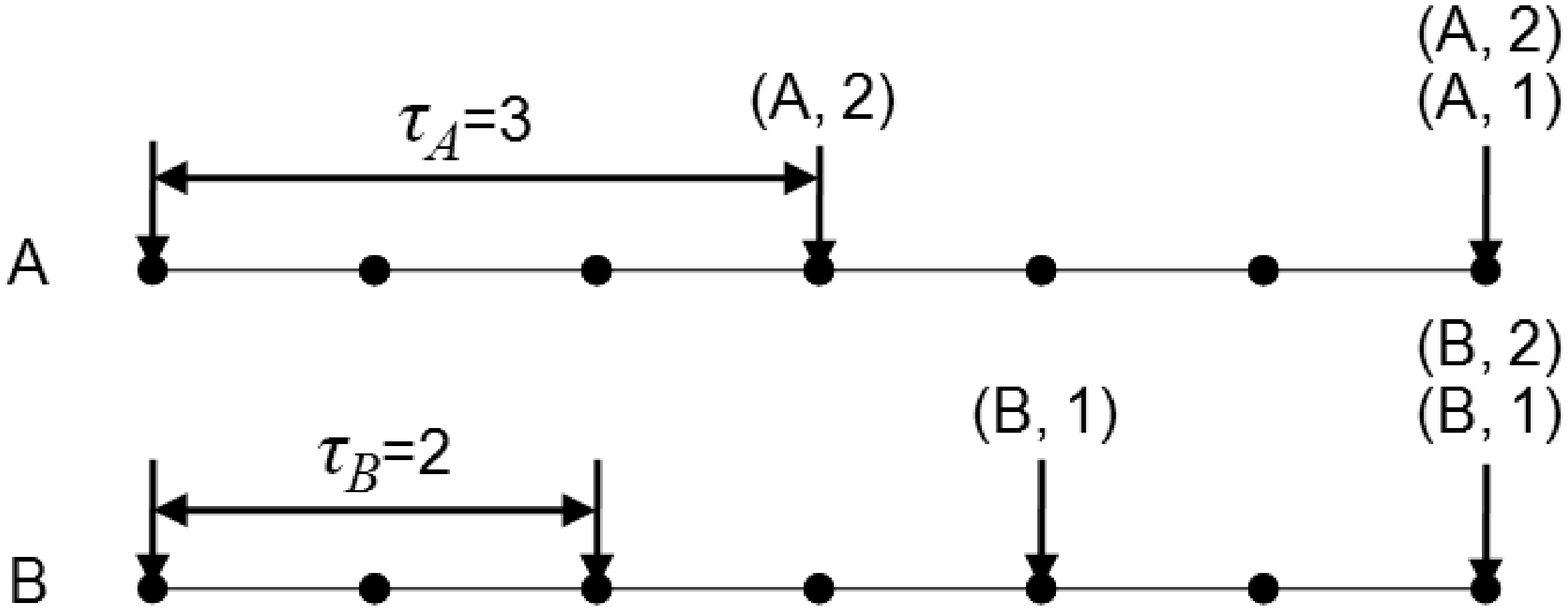}}

\subfloat[The actions scheduled before renumbering]{
\label{fig:feasibility:edf1} 
\includegraphics[width=3.2in]{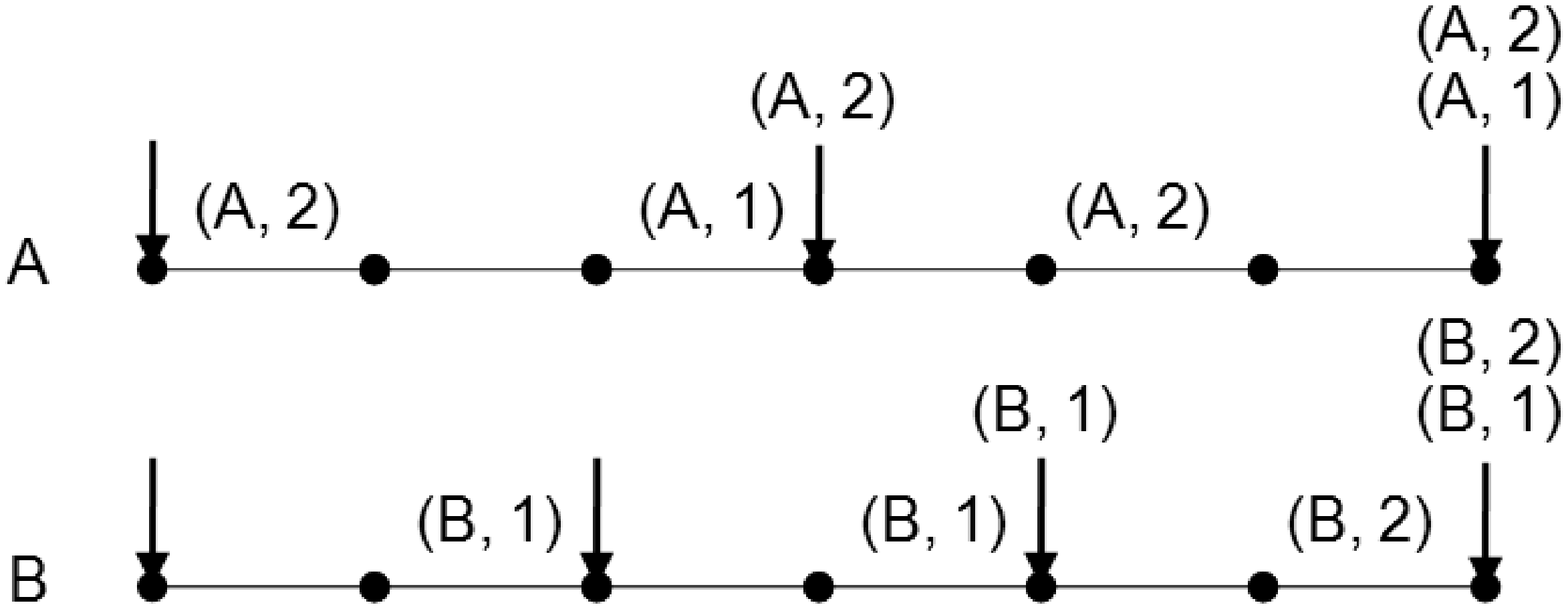}}

\subfloat[The actions scheduled after renumbering]{
\label{fig:feasibility:edf2} 
\includegraphics[width=3.2in]{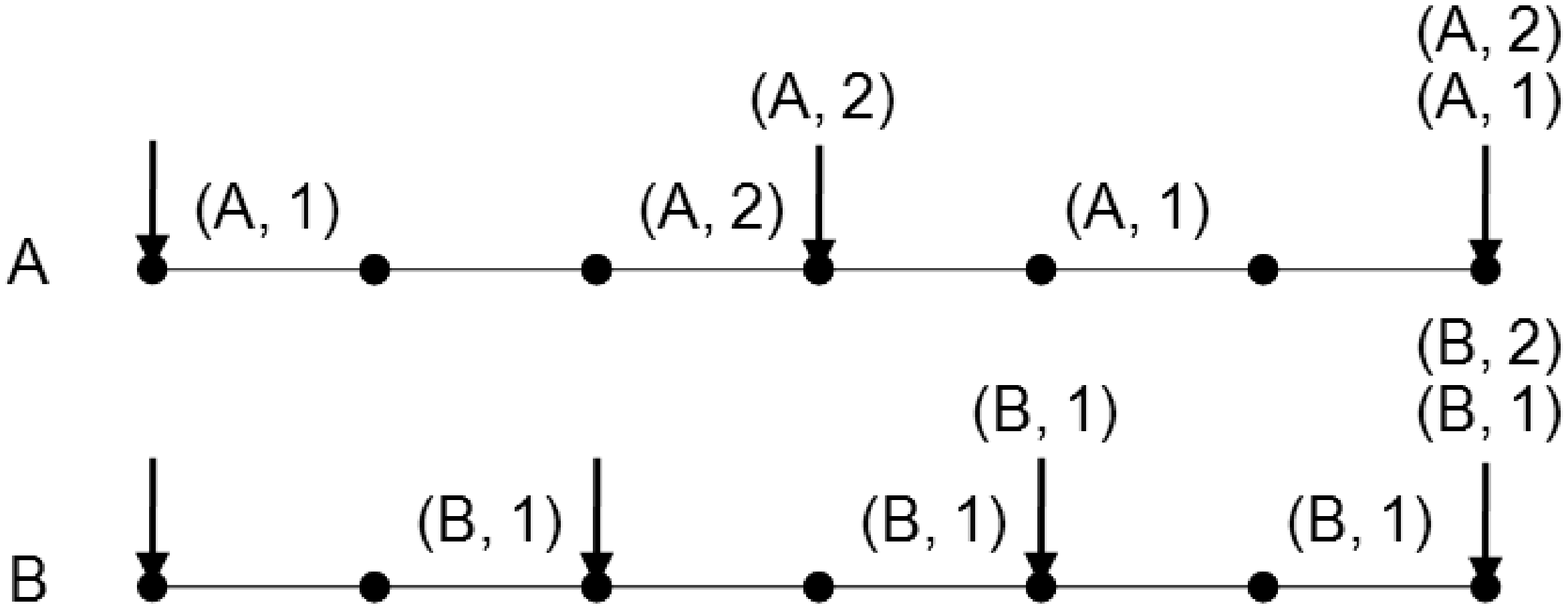}}

\caption{An example of the scheduling policy in Theorem
\ref{theorem:feasibility:integer feasible}. Deadlines of actions are
marked by putting the action above the arrow of its
deadline.}\label{fig:feasibility:edf}
\end{figure}

It remains to show that none of the actions miss their deadlines
under this policy. We prove this by contradiction. Let $d_X(t)$ be
the number of actions of task $X$ whose deadlines are smaller or
equal to $t$. By the way that we mark deadlines of actions, we have
that
$\frac{d_X(\tau_X)}{\tau_X}\leq\frac{d_X(2\tau_X)}{2\tau_X}\leq\dots\leq\frac{d_X(T)}{T}$.
We also have $d_X(t) = d_X(k\tau_X)$, for all $t\in [k\tau_X,
(k+1)\tau_X)$. For any $t$, let
$k_X=\lfloor\frac{t}{\tau_X}\rfloor$, and we then have
\begin{equation}    \label{equation:feasibility:load bound}
\begin{array}{rl} \sum_{X\in S} \frac{d_X(t)}{t} &= \sum_{X\in S}
\frac{d_X(k_X\tau_X)}{t}\\
&\leq \sum_{X\in S} \frac{d_X(k_X\tau_X)}{k_X\tau_X}\\
&\leq
\frac{\sum_{X\in S}d_X(T)}{T}\leq 1,
\end{array}
\end{equation}
where the last inequality follows by condition
(\ref{equation:feasibility:necessary3}).

Suppose there is an action $(X,i)$ that misses its deadline at time
$t$ under our policy. We first consider the case where the policy
schedules an action with deadline smaller or equal to $t$ in all
time slots between 0 and $t$, with no time slot left idle. In this
case, we have $\sum_{X\in S} d_X(t)\geq t+1$ and thus $\sum_{X\in
S}\frac{d_X(t)}{t} > 1$, which contradicts Eq.
(\ref{equation:feasibility:load bound}).

Next we consider the case where at some time $t'<t$ the policy does
not schedule an action with deadline smaller or equal to $t$. That
is, at time $t'$, the policy either schedules an action whose
deadline is strictly larger than $t$ or stays idle. Now we first
claim that $t$ and $t'$ cannot belong to the same period of $X$,
that is, the case with $t'> t-\tau_X$ is not possible. The only
reason that $(X,i)$ is not in fact scheduled at $t'$ is that there
is already one identical action in the corresponding period
containing $t'$. This other action would have a deadline at either
$t$ or $t-k\tau_X$ for some $k\geq 1$. The former case is not
possible because two identical actions cannot have the same
deadline. The latter is also not possible because no action is
scheduled after its deadline and $t-k\tau_X<t'$ if $t$ and $t'$
belong to the same period.

As shown above, the interval $[t-\tau_X+1, t]$ has the property that
all actions scheduled in this interval have deadlines smaller or
equal to $t$. Now we proceed to show that $t'\leq t-\tau_X$ is also
not possible. We do this by expanding this interval while preserving
this property. Pick any action, $(X',i')$, scheduled at time
$t_{X'}$ in the interval and assume that the period of $X'$
containing $t_{X'}$ is $[t^1_{X'},t^2_{X'}]$. We have $t^2_{X'}\leq
t$ since the deadline of $({X'},i')$ is no larger than $t$ and the
deadline of this action is at the end of some period of $X'$. Now,
by the design of the scheduling policy, for any $\bar{t}_{X'}$ in
$[t^1_{X'},t_{X'}]$, the action scheduled in $\bar{t}_{X'}$ should
have deadline smaller or equal to the deadline of $({X'},i')$, or
otherwise $({X'},i')$ should have been scheduled in $\bar{t}_{X'}$.
The deadline of the action scheduled in $\bar{t}_{X'}$ is thus also
smaller or equal to $t$. Thus, if $t^1_{X'}$ is smaller than the
beginning of the interval, we can expand the interval to
$[t^1_{X'},t]$ while preserving the desired property. We keep
expanding the interval until no more expansions are possible.

Let $(Y,j)$ be the action in the resulting interval with the largest
period, $\tau_Y$, and suppose that it is scheduled at time $t_Y$.
Assume that the period of $Y$ containing $t_Y$ is $[t^1_Y, t^2_Y]$.
By the way we expand the interval, $[t^1_Y, t^2_Y]$ is within the
expanded interval and all actions scheduled in $[t^1_Y, t^2_Y]$ have
deadlines smaller or equal to $t$. For each action in $[t^1_Y,
t^2_Y]$, the reason that it has not been scheduled earlier at time
$t'$ is because there is already one identical action scheduled in
its period that contains $t'$. This identical action, also with
deadline earlier than $t$, must have been scheduled before time
$t'$. Since the period of this action is smaller or equal to
$\tau_Y$, its identical counterpart must have been scheduled in
$[t'-\tau_Y+1, t'-1]$. However, there are at most $\tau_Y-1$ actions
scheduled in $[t'-\tau_Y+1, t'-1]$, while there are $\tau_Y$ actions
in $[t^1_Y, t^2_Y]$, leading to a contradiction. Thus, this case is
also not possible.

In sum, all actions are scheduled before their deadlines using this
policy, and the proof is completed.
\end{IEEEproof}

Now we can derive the necessary and sufficient condition for a
system to be feasible.

\begin{theorem}   \label{theorem:feasibility:sufficient}
A system with a set of tasks $S=\{A,B,\dots\}$ is feasible if and
only if there exists $\{f^i_X|X\in S, 1\leq i\leq \tau_X\}$ such
that (\ref{equation:feasibility:necessary1}) -
(\ref{equation:feasibility:necessary3}) are satisfied.
\end{theorem}
\begin{IEEEproof}
Lemma \ref{lemma:feasibility:necessary} has established that these
conditions are necessary. It remains to show that they are also
sufficient. Suppose there exists $f = \{f^i_X|X\in S, 1\leq i\leq
\tau_X\}$ that satisfy (\ref{equation:feasibility:necessary1}) -
(\ref{equation:feasibility:necessary3}), then Theorem
\ref{theorem:feasibility:integer hull} shows that there exists
integer vectors $n[1], n[2],\dots, n[v]$ such that $f=\sum_{u=1}^v
\alpha_u n[u]$, where $\alpha_u$'s are positive numbers with
$\sum_{u=1}^v \alpha_u=1$. Let $\eta_u$ be the scheduling policy for
the integer vector $n[u]$ as in the proof of Theorem
\ref{theorem:feasibility:integer feasible}, for each $u$. Theorem
\ref{theorem:feasibility:integer feasible} have shown that for each
$u$, the average reward obtained by $X$ under $\eta_u$,
$q_X[\eta_u]$, is at least $\sum_{i=1}^{\tau_X} n[u]^i_Xr^i_X$.
Finally, we can design a policy as a weighted round robin policy
that switches among the policies $\eta_1, \eta_2,\dots,\eta_v$ ,
with policy $\eta_u$ being chosen in $\alpha_u$ of the frames. The
average reward obtained by $X$ is hence $q_X=\sum_u
\alpha_uq_X[\eta_u]\geq\sum_u \alpha_u(\sum_{i=1}^{\tau_X}
n[u]^i_Xr^i_X)=\sum_{i=1}^{\tau_X}f^i_Xr^i_X\geq q^*_X$. Thus, this
policy fulfills the system and so the conditions are also
sufficient.
\end{IEEEproof}

Using Theorem \ref{theorem:feasibility:sufficient}, checking whether
a system is feasible can be done by any linear programming solver.
The computational overhead for checking feasibility can be further
reduced by using the fact that $r^{i}_X\geq r^{j}_X$, for all $i<j$.
Given a system and $\{f^i_X\}$ that satisfies conditions
(\ref{equation:feasibility:necessary1}) -
(\ref{equation:feasibility:necessary3}) with
$f^j_Y<\frac{T}{\tau_Y}$ and $f^k_Y>0$ for some $j<k$ and $Y\in S$.
Let $\delta=\min\{\frac{T}{\tau_Y}-f^j_Y, f^k_Y\}$. Construct
$\{\hat{f}^i_X\}$ such that $\hat{f}^j_Y = f^j_Y+\delta$,
$\hat{f}^k_Y = f^k_Y-\delta$, and $\hat{f}^i_X = f^i_X$ for all
other elements. Then $\{\hat{f}^i_X\}$ also satisfies conditions
(\ref{equation:feasibility:necessary1}) -
(\ref{equation:feasibility:necessary3}). Based on this observation,
we derive an algorithm for checking feasibility as shown in
Algorithm \ref{algorithm:feasibility:check}. This essentially
transfers slots from less reward earning actions to more reward
earning actions. The running time of this algorithm is $O(\sum_{X\in
S}\tau_X)$. Since a specification of a system involves at least the
$\sum_{X\in S}\tau_X$ variables of $\{r^i_X\}$, Algorithm
\ref{algorithm:feasibility:check} is essentially a linear time
algorithm.

\begin{algorithm}[h]
\caption{Feasibility Checker} \label{algorithm:feasibility:check}
\begin{algorithmic}[1]
\REQUIRE $S$, $\{\tau_X|X\in S\}$, $\{r^i_X|X\in S, 1\leq i\leq
\tau_X\}$, $\{q^*_X|X\in S\}$

\FOR{$X\in S$} \IF{$q^*_X > \frac{T}{\tau_X}\sum_{i=1}^{\tau_X}
r^i_X$}\RETURN Infeasible\ENDIF\ENDFOR

\FOR{$X\in S$}

\STATE $i\leftarrow 1$ \\
\WHILE{$q^*_X>0$}

\IF{$q^*_X > \frac{T}{\tau_X}r^i_X$}
\STATE $f^i_X\leftarrow T/\tau_X$\\
\STATE $q^*_X \leftarrow q^*_X-\frac{T}{\tau_X}r^i_X$\\
\ELSE
\STATE $f^i_X\leftarrow q^*_X/r^i_X$\\
\STATE $q^*_X\leftarrow 0$\\
\ENDIF
\STATE $i\leftarrow i+1$\\

\ENDWHILE \ENDFOR

\IF{$\sum_{X\in S}\sum^{i=1}_{\tau_X} f^i_X \leq T$} \RETURN
Feasible\\
\ELSE \RETURN Infeasible\\
\ENDIF

\end{algorithmic}
\end{algorithm}

In addition to evaluating feasibility, the proof of Theorem
\ref{theorem:feasibility:sufficient} also demonstrates an off-line
feasibility optimal policy. In many scenarios, however, on-line
policies are preferred. In the next section, we introduce a
guideline for designing scheduling policies that turns out to
suggest simple on-line policies.

\section{Designing Scheduling Policies}
\label{section:scheduling}

In this section, we study the problem of designing scheduling
policies. We establish sufficient conditions for a policy to be
either feasibility optimal or $p$-approximately so.

We start by introducing a metric to evaluate the performance of a
policy $\eta$. Let $\tilde{q}_X(k)$ be the total reward obtained by
task $X$ during the frame $((k-1)T, kT]$. We then have
$q_X=\liminf_{k\rightarrow\infty}\frac{\sum_{i=1}^k\tilde{q}_X(i)}{k}$.
We also define the \emph{debt} of task $X$.
\begin{definition}
The \emph{debt} of task $X$ in the frame $((k-1)T, kT]$, $d_X(k)$ is
defined recursively as follows:
\begin{align*}
&d_X(0) = 0,\\
&d_X(k) = [d_X(k-1)+q^*_X-\tilde{q}_X(k)]^+, \forall k>0.
\end{align*}
\end{definition}

\begin{lemma}
A system is fulfilled by a policy $\eta$ if
$\lim_{k\rightarrow\infty} d_X(k)/k=0$ with probability 1.
\end{lemma}
\begin{IEEEproof}
We have $d_X(k)\geq kq^*_X-\sum_{i=1}^k\tilde{q}_X(i)$ and
$d_X(k)/k\geq q^*_X-\frac{1}{k}\sum_{i=1}^k\tilde{q}_X(i)$. Thus, if
$\lim_{k\rightarrow\infty} d_X(k)/k=0$, then
$q_X=\liminf_{k\rightarrow\infty}\frac{\sum_{i=1}^k\tilde{q}_X(i)}{k}\geq
q^*_X$ and the system is fulfilled.
\end{IEEEproof}

We can describe the \emph{state} of the system in the $k^{th}$ frame
by the debts of tasks, $[d_X(k)|X\in S]$. Consider a policy that
schedules jobs solely based on the requirements and the state of the
system. The evolution of the state of the system can then be
described as a Markov chain.
\begin{lemma}\label{lemma:scheduling:positive recurrent}
Suppose the evolution of the state of a system can be described as a
Markov chain under some policy $\eta$. The system is fulfilled by
$\eta$ if this Markov chain is irreducible and positive recurrent.
\end{lemma}
\begin{IEEEproof}
Since the Markov chain is positive recurrent, the state
$\{d_X(k)=0,\forall X\in S\}$ is visited infinitely many times.
Further, assuming that the system is in this state at frames
$k_1,k_2,k_3,\dots$, then $[k_{n+1}-k_n]$ is a series of i.i.d.
random variables with finite mean. Let $I_n$ be the indicator
variable that there exists some $\tilde{k}_n$ between frame $k_n$
and frame $k_{n+1}$ such that $d_X(\tilde{k}_n)/\tilde{k}_n> \delta$
for some $X\in S$, for some arbitrary $\delta>0$. Let
$q^*_{max}=\max_{X\in S}{q^*_X}$. If $I_n=1$, we have that
$\tilde{k}_n\geq k_n\geq n$ and thus $d_X(\tilde{k}_n)>n\delta$, for
some $X\in S$. Since $d_X(k)$ can be incremented by at most
$q^*_{max}$ in a frame and $d_X(k_n)=0$,
$\tilde{k}_n-k_n>n\delta/q^*_{max}$ and
$k_{n+1}-k_n>\tilde{k}_n-k_n>n\delta/q^*_{max}$. Thus,
\begin{align*}
Prob\{I_n=1\}&<Prob\{k_{n+1}-k_n>n\delta/q^*_{max}\}\\
&=Prob\{k_{2}-k_1>n\delta/q^*_{max}\},
\end{align*}
and
\begin{align*}
\sum_{n=1}^\infty Prob\{I_n=1\}&< \sum_{n=1}^\infty
Prob\{k_{2}-k_1>n\delta/q^*_{max}\}\\
&\leq E[k_2-k_1]<\infty,
\end{align*}
By Borel-Cantelli Lemma, the probability that $I_n=1$ for infinitely
many $n$'s is zero, and so is the probability that $d_X(k)/k>\delta$
for infinitely many $k$'s. Thus,
$\limsup_{k\rightarrow\infty}d_X(k)/k < \delta$ with probability 1,
for all $X\in S$ and any arbitrary $\delta>0$. Finally, we have
$\lim_{k\rightarrow\infty} d_X(k)/k=0$ with probability 1 since
$d_X(k)\geq 0$ by definition.
\end{IEEEproof}

Based on the above lemmas, we determine a sufficient condition for a
policy to be a $p$-approximation policy. The proof is based on the
Foster-Lyapunov Theorem:

\begin{theorem}[Foster-Lyapunov Theorem]\label{theorem:scheduling:lyapunov}
Consider a Markov chain with state space $\mathcal{D}$. Let $D(k)$
be the state of the Markov chain at the $k^{th}$ step. If there
exists a non-negative function $L:\mathcal{D}\rightarrow R$, a
positive number $\delta$, and a finite subset $\mathcal{D}_0$ of
$\mathcal{D}$ such that:
\begin{align*}
E[L(D(k+1))-L(D(k))|D(k)] \leq -\delta,&\mbox{ if $D(k)\notin
\mathcal{D}_0$},\\
E[L(D(k+1))|D(k)]<\infty, &\mbox{ if $D(k)\in \mathcal{D}_0$},
\end{align*}
then the Markov chain is positive recurrent.      $\blacksquare$
\end{theorem}

\begin{theorem} \label{theorem:scheduling:approximation}
A policy $\eta$ is a $p$-approximation policy, for some $p>1$, if it
schedules jobs solely based on the requirements and the state of a
system and, for each $k$, the following holds:
\[
\sum_{X\in S} \tilde{q}_X(k)d_X(k)\geq (\max_{\mbox{$[q_X]$: $[q_X]$
is feasible}}\sum_{X\in S} q_Xd_X(k))/p.
\]
\end{theorem}
\begin{IEEEproof}
Consider a system with minimum reward requirements $[q^*_X]$ such
that the same system with minimum reward requirements $[pq^*_X]$ is
also strictly feasible. By Lemma \ref{lemma:scheduling:positive
recurrent}, it suffices to show that under the policy $\eta$, the
resulting Markov chain is positive recurrent. Consider the Lyapunov
function $L(k):=\sum_{X\in S} d^2_X(k)/2$. The Lyapunov drift
function can be written as:
\begin{align*}
\Delta L(k+1) :=& L(k+1)-L(k) = \frac{1}{2}\sum_{X\in S} [d^2_X(k+1) - d^2_X(k)]\\
\leq &\sum_{X\in S} (q^*_X-\tilde{q}_X(k))d_X(k)+C,
\end{align*}
where $C$ is a bounded constant. Since $[pq^*_X]$ is also strictly
feasible and $d_X(k)\geq 0$, there exists $\epsilon>0$ such that
$(1+\epsilon)\sum_{X\in S} pq^*_Xd_X(k)\leq \max_{\mbox{$[q_X]$:
$[q_X]$ is feasible}} \sum_{X\in S} q_Xd_X(k)$, and hence
$(1+\epsilon)\sum_{X\in S} q^*_Xd_X(k)\leq \sum_{X\in S}
\tilde{q}_X(k)d_X(k)$. Thus, we have
\begin{equation}
\Delta L(k+1)\leq -\epsilon\sum_{X\in S} q^*_Xd_X(k)+C.
\end{equation}
Let $\mathcal{D}_0$ be the set of states $[d_X(k)|X\in S]$ with
$\sum_{X\in S} q^*_Xd_X(k) < (C+\delta)/\epsilon$, for some positive
finite number $\delta$. Then, $\mathcal{D}_0$ is a finite set (since
$q^*_X>0$ for all $x\in S$), with $\Delta L(k+1)<-\delta$ when the
state of frame $k$ is not in $\mathcal{D}_0$. Further, since
$d_X(k)$ can be increased by at most $q^*_X$ in each frame, $L(k+1)$
is finite if the state of frame $k$ is in $\mathcal{D}_0$. By
Theorem \ref{theorem:scheduling:lyapunov}, this Markov chain is
positive recurrent and policy $\eta$ fulfills this system.
\end{IEEEproof}

Since a $1$-approximation policy is also a feasibility optimal one,
a similar proof yields the following:
\begin{theorem} \label{theorem:scheudling:optimal}
A policy $\eta$ fulfills a strictly feasible system if it maximizes
$\sum_{X\in S} \tilde{q}_X(k)d_X(k)$ among all feasible $[q_X]$ in
every frame $k$. It is a feasibility optimal policy if the above
holds for all strictly feasible systems.
\end{theorem}

\section{An On-Line Scheduling Policy}
\label{section:greedy}

While Section \ref{section:scheduling} has described a sufficient
condition for designing feasibility optimal policies, the overhead
for computing such a feasibility optimal scheduling policy may be
too high to implement. In this section, we introduce a simple
on-line policy. We also analyze the performance of this policy under
different scenarios.

Theorem \ref{theorem:scheudling:optimal} has shown that a policy
that maximizes $\sum_{X\in S} \tilde{q}_X(k)d_X(k)$ among all
feasible $[q_X]$ in every frame $k$ is feasibility optimal. The
on-line policy follows this guideline by greedily selecting the job
with the highest $r^i_xd_X(k)$ in each time slot. Assume that, at
some time $t$ in frame $k$, task $X$ has already been scheduled
$i_X$ times in its period. The on-line policy then schedules the
task $Y$ so that $r^{i_Y+1}_Yd_Y(k)$ is maximized among all $X\in
S$. A more detailed description of this policy, which we call the
\emph{Greedy Maximizer}, is shown in Algorithm
\ref{algorithm:greedy:greedy}.

\begin{algorithm}[h]
\caption{Greedy Maximizer} \label{algorithm:greedy:greedy}
\begin{algorithmic}[1]
\REQUIRE $S$, $\{\tau_X|X\in S\}$, $\{r^i_X|X\in S, 1\leq i\leq
\tau_X\}$, $\{q^*_X|X\in S\}$
\STATE $T\leftarrow$ least common multiplier of $\{\tau_X|X\in S\}$\\
\FOR{$X\in S$}\STATE $d_X\leftarrow 0$\ENDFOR
\STATE $k\leftarrow 0$\\
\STATE $t\leftarrow 0$\\
\LOOP

\STATE $t\leftarrow t+1$\\
\IF{$t\mod T=1$\COMMENT{A new frame}}

\STATE $k\leftarrow k+1$\\
\FOR{$X\in S$}\STATE $d_X\leftarrow
[d_X+q^*_X-\tilde{q}_X]^+$\\
\STATE $\tilde{q}_X\leftarrow 0$\\
\ENDFOR

\ENDIF

\FOR{$X\in S$}\IF{$t\mod \tau_X=1$\COMMENT{A new period for $X$}}

\STATE $i_X\leftarrow 0$

\ENDIF\ENDFOR

\STATE $Y\leftarrow \arg\max_{X\in S}r^{i_X+1}_Xd_X$\\
\STATE $i_Y\leftarrow i_Y+1$\\
\STATE $\tilde{q}_Y\leftarrow \tilde{q}_Y+r^{i_Y}_Y$\\
\STATE execute the job of $Y$ at time $t$\\

\ENDLOOP
\end{algorithmic}
\end{algorithm}

Next, we evaluate the performance of the Greedy Maximizer. We show
that this policy is feasibility optimal if the periods of all tasks
are the same, and that it is $2$-approximation in general.

\begin{theorem} \label{theorem:greedy:opt}
The Greedy Maximizer fulfills all strictly feasible systems with
$\tau_X\equiv\tau$, for all $X\in S$.
\end{theorem}
\begin{IEEEproof}
It suffices to prove that the Greedy Maximizer indeed maximizes
$\sum_{X\in S}d_X(k)\tilde{q}_X(k)$ in every frame. Suppose at some
frame $k$, the debts are $\{d_X(k)\}$ and the schedule generated by
the Greedy Maximizer is $\eta_{GM}(t)$, $t\in (kT,(k+1)T]$. Let
$GM:=\sum_{X\in S}d_X(k)\tilde{q}_X(k)$ when $\eta_{GM}$ is applied.
Consider another schedule, $\eta_{OPT}(t)$, that achieves
Max$\sum_{X\in S}d_X(k)\tilde{q}_X(k)=:OPT$ in this frame. We need
to show that $GM\geq OPT$.

We are going to modify $\eta_{OPT}(t)$ slot by slot until it is the
same as $\eta_{GM}$. Let $\eta^{i}_{OPT}(t)$ be the schedule after
we have made sure $\eta^{i}_{OPT}(t)=\eta_{GM}(t)$ for all $t$
between $kT$ and $i$, and let $OPT^{i}:=\sum_{X\in
S}d_X(k)\tilde{q}_X(k)$ when $\eta^i_{OPT}(t)$ is applied. We then
have $\eta_{OPT}\equiv\eta^{kT}_{OPT}$ and
$\eta_{GM}\equiv\eta^{(k+1)T}_{OPT}$. The process of modification is
as follows: If $\eta^{i}_{OPT}(i+1)=\eta_{GM}(i+1)$, then we do not
need to modify anything and we simply set $\eta^{i+1}_{OPT}\equiv
\eta^{i}_{OPT}$. On the other hand, if $\eta^{i}_{OPT}(i+1)\neq
\eta_{GM}(i+1)$, say, $\eta_{GM}(i+1)=(A,j_A)$ and
$\eta^{i}_{OPT}(i+1)=(B,j_B)$, then we modify $\eta^{i}_{OPT}(t)$
under two different cases. The first case is that $\eta^{i}_{OPT}$
is going to schedule the action $(A,j_A)$ some time after $i+1$ in
this frame. In this case $\eta^{i+1}_{OPT}$ is obtained by switching
the two actions $(A,j_A)$ and $(B,j_B)$ in $\eta^{i}_{OPT}$. One
such example is shown in Fig. \ref{fig:greedy:opt1}. Since
interchanging the order of actions does not influence the value of
$\sum_{X\in S}d_X(k)\tilde{q}_X(k)$, we have $OPT^{i+1}=OPT^{i}$ for
this case. The second case is that $\eta^{i}_{OPT}$ does not
schedule the action $(A,j_A)$ in the frame. Then $\eta^{i+1}_{OPT}$
is obtained by setting $\eta^{i+1}_{OPT}(i+1)=(A,j_A)$ and
scheduling the same jobs as $\eta^{i}_{OPT}$ for all succeeding time
slots. Since the Greedy Maximizer schedules $(A,j_A)$ in this slot,
we have $r^{j_A}_Ad_A(k)\geq r^{j_B}_Bd_B(k)$. Also, for all
succeeding time slots, if job $B$ is scheduled, then the reward for
that slot is going to be increased since the number of executions of
job $B$ has been decreased by 1; if a job $C$ other than $A$ and $B$
is scheduled, then the reward for that slot is not influenced by the
modification. Fig. \ref{fig:greedy:opt2} has illustrated one such
example. In sum, we have that $OPT^{i+1}\geq OPT^{i}$.

We have established that $OPT^{i+1}\geq OPT^{i}$ for all $i\in[kT,
(k+1)T]$. Since $OPT=OPT^{kT}$ and $GM=OPT^{(k+1)T}$, we have
$GM\geq OPT$ and thus the Greedy Maximizer indeed maximizes
$\sum_{X\in S}d_X(k)\tilde{q}_X(k)$.

\begin{figure}[t]\subfloat[First case]{
\label{fig:greedy:opt1} 
\includegraphics[width=3.2in]{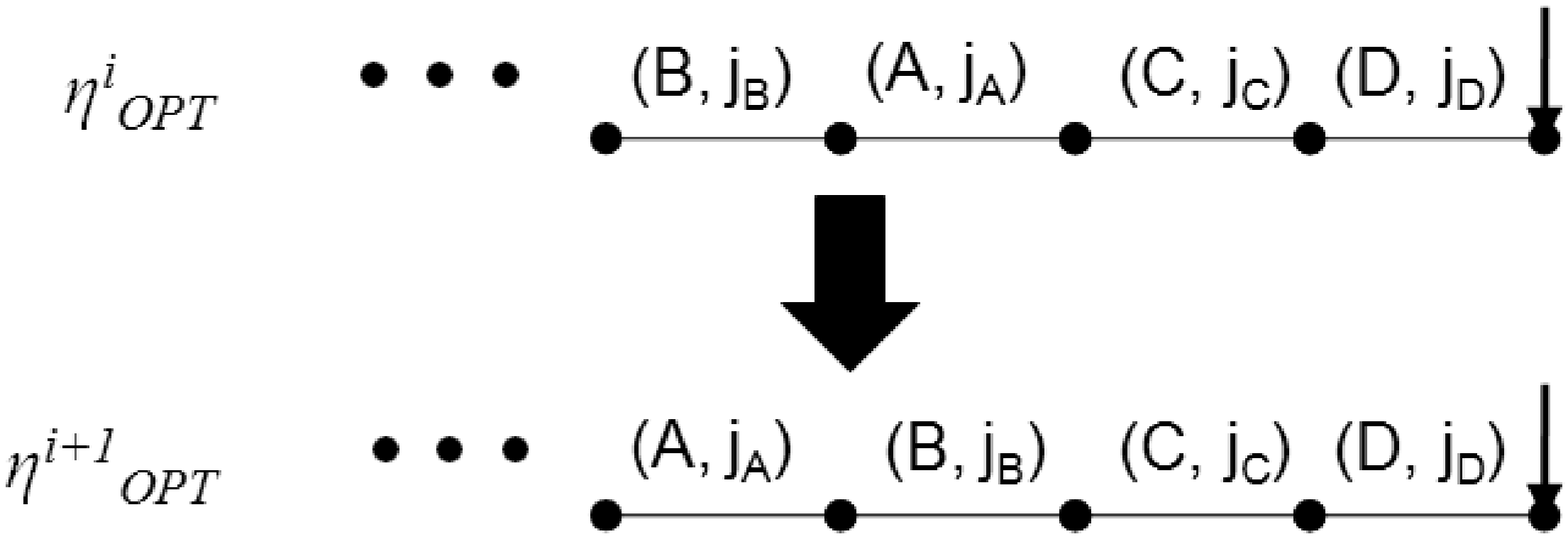}}

\subfloat[Second case]{
\label{fig:greedy:opt2} 
\includegraphics[width=3.2in]{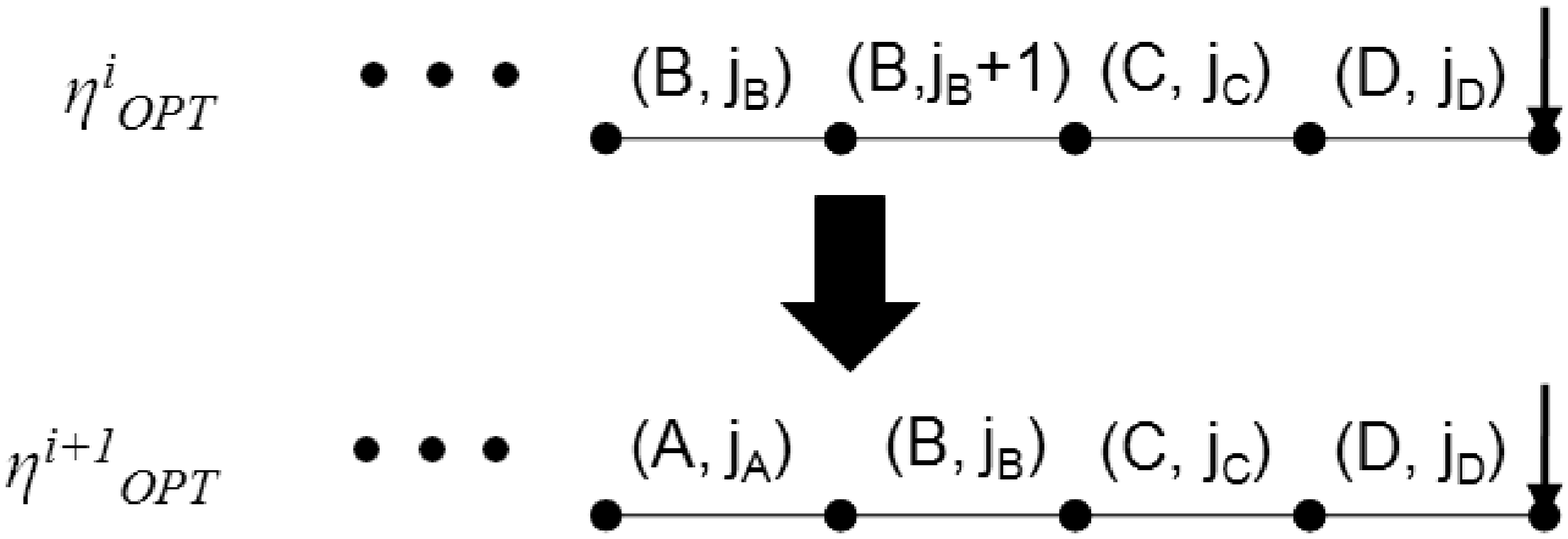}}

\caption{Examples of modification in Theorem
\ref{theorem:greedy:opt}}\label{fig:greedy:greedy_opt}
\end{figure}

\end{IEEEproof}

However, when the periods of tasks are not the same, the Greedy
Maximizer does not always maximize $\sum_{X\in
S}d_X(k)\tilde{q}_X(k)$ and thus may not be feasibility optimal. An
example is given below.

\begin{figure}[t]\subfloat[Greedy Maximizer]{
\label{fig:greedy:gm1} 
\includegraphics[width=3.2in]{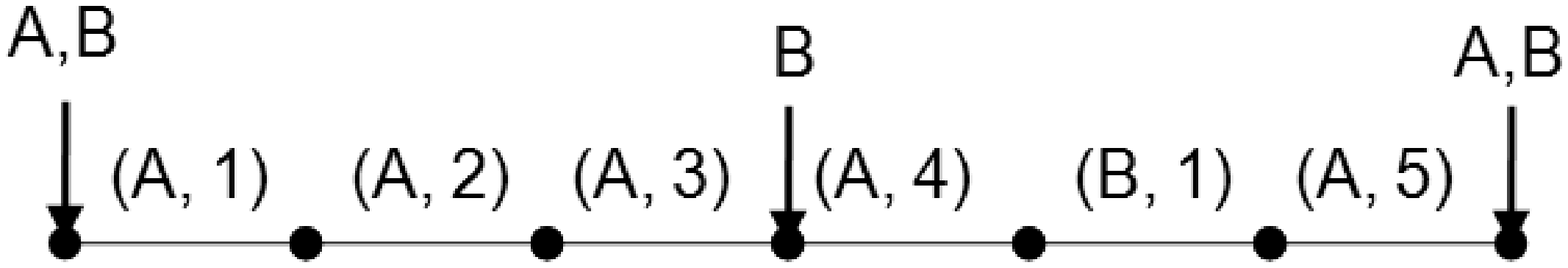}}

\subfloat[Feasibility optimal scheduler]{
\label{fig:greedy:gm2} 
\includegraphics[width=3.2in]{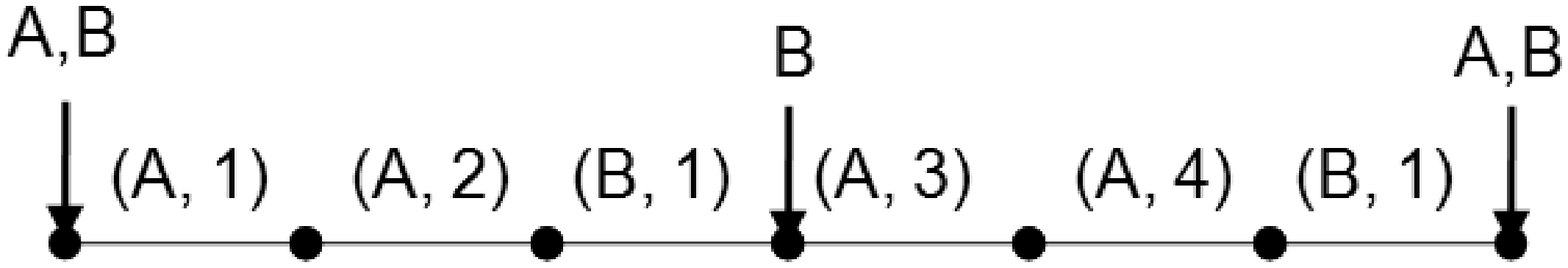}}

\caption{An example of the resulting schedule by the Greedy
Maximizer and a feasibility optimal scheduler, respectively in
Example \ref{example:example}.}\end{figure}

\begin{example} \label{example:example}
Consider a system with two tasks, $A$ and $B$, with $\tau_A=6$,
$\tau_B=3$. Assume that $r^1_A=r^2_A=r^3_A=r^4_A=100$,
$r^5_A=r^6_A=1$, $r^1_B=10$, and $r^2_B=r^3_B=0$. Suppose, at some
frame $k$, $d_A(k)=d_B(k)=1$. The Greedy Maximizer would schedule
jobs as in Fig. \ref{fig:greedy:gm1}, and yield
$d_A(k)\tilde{q}_A(k)+d_B(k)\tilde{q}_B(k)=411$. On the other hand,
a feasibility optimal scheduler would schedule jobs as in Fig.
\ref{fig:greedy:gm2}, and yield
$d_A(k)\tilde{q}_A(k)+d_B(k)\tilde{q}_B(k)=420$.  $\blacksquare$
\end{example}

Although the Greedy Maximizer is not feasibility optimal, we can
still derive an approximation bound for this policy.

\begin{figure}[t]\subfloat[First case]{
\label{fig:greedy:approx1} 
\includegraphics[width=3.2in]{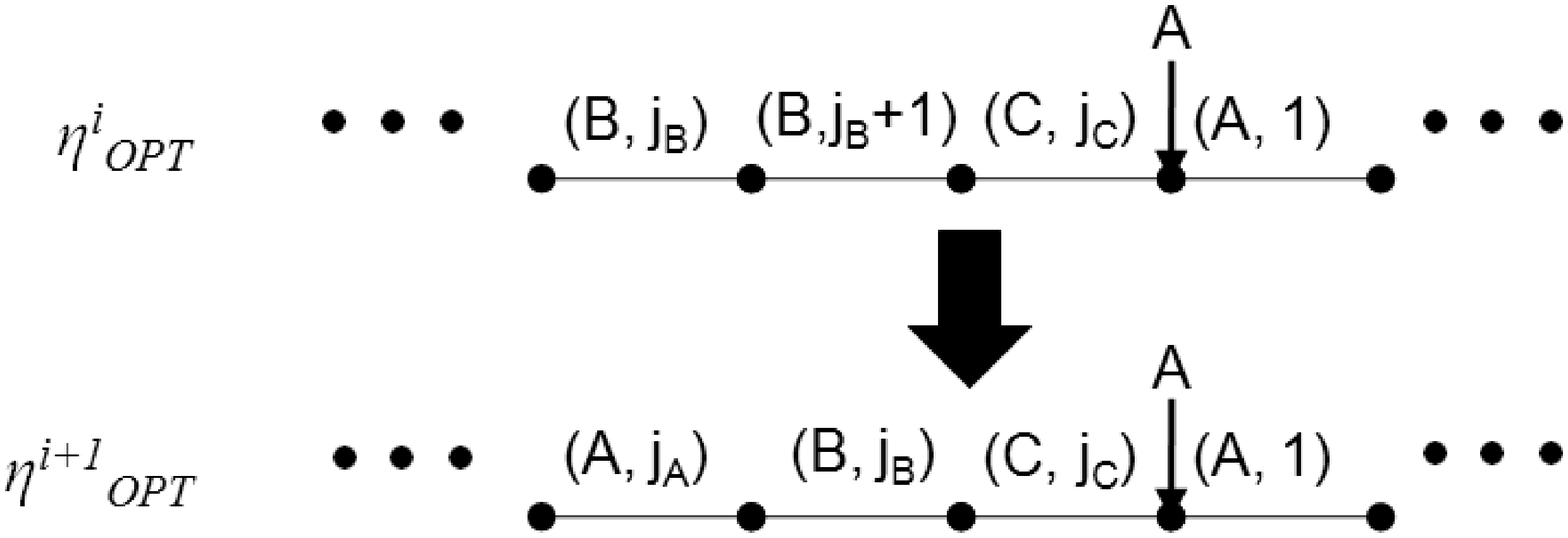}}

\subfloat[Second case]{
\label{fig:greedy:approx2} 
\includegraphics[width=3.2in]{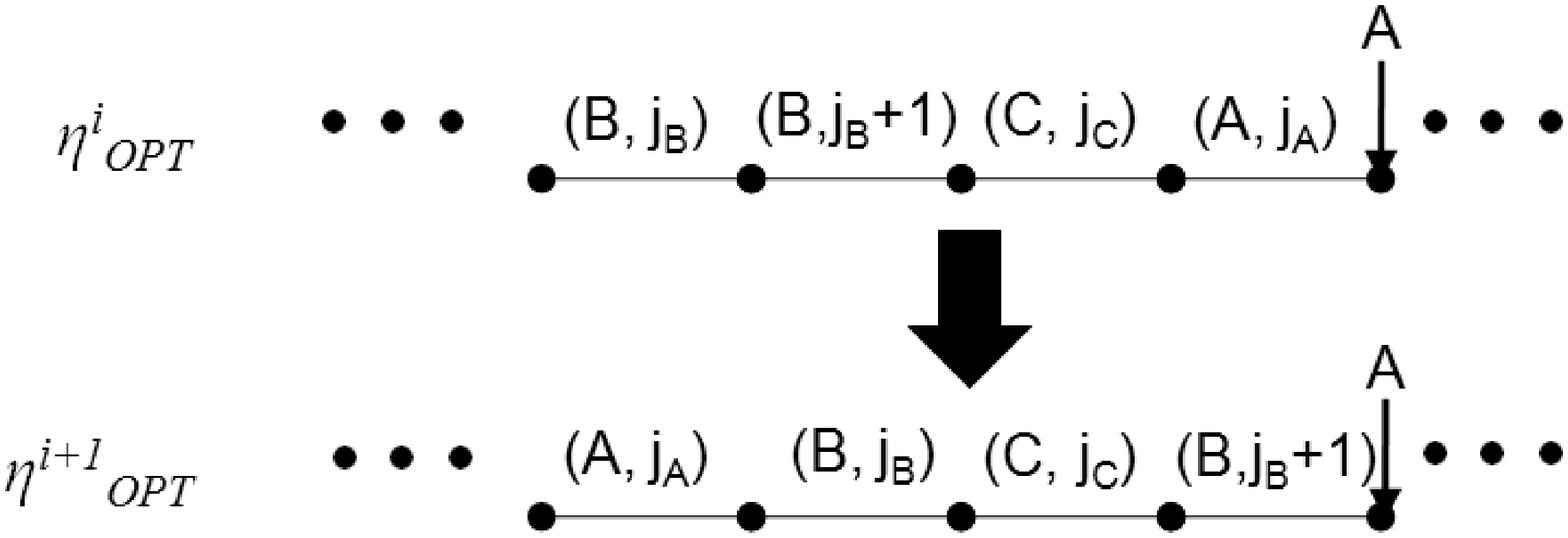}}

\subfloat[Third case]{
\label{fig:greedy:approx3} 
\includegraphics[width=3.2in]{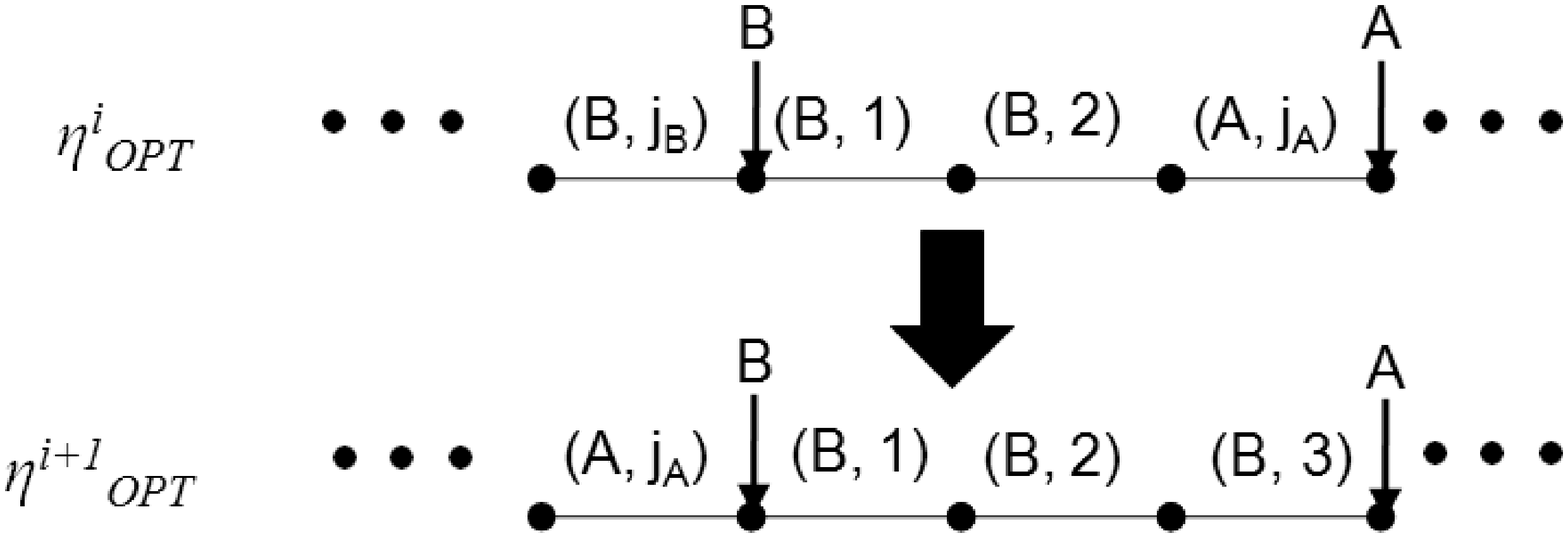}}

\caption{Examples of modification in Theorem
\ref{theorem:greedy:approx}}\label{fig:greedy:greedy_approx}
\end{figure}

\begin{theorem} \label{theorem:greedy:approx}
The Greedy Maximizer is a $2$-approximation policy.
\end{theorem}
\begin{IEEEproof}
The proof is similar to that of Theorem \ref{theorem:greedy:opt}.
Define $\eta_{GM}, \eta_{OPT}, \eta^{i}_{OPT}, GM, OPT,$ and
$OPT^{i}$ in the same way as in the proof of Theorem
\ref{theorem:greedy:opt}. By Theorem
\ref{theorem:scheduling:approximation}, it suffices to show that
$GM\geq OPT/2$.

We obtain $\eta^{i}_{OPT}$ as follows: If
$\eta_{GM}(i+1)=\eta^{i}_{OPT}(i+1)$, then we set
$\eta^{i+1}_{OPT}\equiv\eta^{i}_{OPT}$. If
$\eta_{GM}(i+1)=(A,j_A)\neq (B,j_B)=\eta^{i}_{OPT}(i)$, then we
consider three possible cases. The first case is that the job
$(A,j_A)$ is not scheduled by $\eta^{i}_{OPT}$ in this period of
$A$. In this case, we set $\eta^{i+1}_{OPT}(i+1)=(A,j_A)$ and use
the same schedule as $\eta^{i}_{OPT}$ for all succeeding time slots.
An example is shown in Fig. \ref{fig:greedy:approx1}. The same
analysis in Theorem \ref{theorem:greedy:opt} shows that
$OPT^{i+1}\geq OPT^{i}$. The second case is that the job $(A,j_A)$
is scheduled by $\eta^{i}_{OPT}$ in this period of $A$ and there is
no deadline of $B$ before the deadline of $A$. In this case, we
obtain $\eta^{i+1}_{OPT}$ by switching the jobs $(A,j_A)$ and
$(B,j_B)$ in $\eta^{i}_{OPT}$. An example is shown in Fig.
\ref{fig:greedy:approx2}. We have $OPT^{i}=OPT^{i+1}$ for this case.
The last case is that the job $(A,j_A)$ is scheduled by
$\eta^{i}_{OPT}$ in this period of $A$, and there is a deadline of
$B$ before the deadline of $A$. In this case also, we obtain
$\eta^{i+1}_{OPT}$ by switching the two jobs and renumbering these
jobs if necessary. The rewards obtained by all tasks other than $B$
are not influenced by this modification. However, as the example
shown in Fig. \ref{fig:greedy:approx3}, the job $(B,j_B)$ in
$\eta^{i}_{OPT}$ may become a job $(B,j'_B)$ in $\eta^{i+1}_{OPT}$
with $j'_B>j_B$. Thus, the reward obtained by $B$ may be decreased.
However, since rewards are non-negative, the amount of loss for $B$
is at most $r^{j_B}_B$. By the design of Greedy Maximizer, we have
$r^{j_A}_Ad_A(k)\geq r^{j_B}_Bd_B(k)$ and thus $OPT^{i+1}\geq
OPT^{i}-r^{j_A}_Ad_A(k)$.

In sum, for all $i$, if the Greedy Maximizer schedules $(A,j_A)$ at
time slot $i+1$, we have $OPT^{i+1}\geq OPT^{i}-r^{j_A}_Ad_A(k)$.
Thus, $GM=OPT^{(k+1)T}\geq OPT^{kT}-GM=OPT-GM$ and $GM\geq OPT/2$.
\end{IEEEproof}

\section{Simulation Results}
\label{section:simulation}

In this section, we present our simulation results. We first
consider a system with six tasks, each with different period,
$\tau_X$, length of mandatory part, $m_X$, and optional part, $o_X$.
Let $f_X(t)$ be the total reward $X$ obtained in a period if it
executes $t$ time slots of its optional part in the period. Thus,
per our model, we have
\begin{equation*}
r^i_X=\left \{ \begin{array}{rl}
M, & \mbox{if $i\leq m_X$,}\\
f_X(i-m_X+1)\\\mbox{   }-f_X(i-m_X), & \mbox{if $m_X < i \leq m_X+o_X$,}\\
0, & \mbox{if $i > m_X+o_X$.}\end{array}\right.
\end{equation*}

As in \cite{HA01}, we consider three different types of function
$f_X$: exponential, logarithmic, and linear. The reward requirement
of $X$ is $\frac{T}{\tau_X}m_XM+\hat{q}^*_X$. We compare the set of
requirements of tasks that can be fulfilled by the Greedy Maximizer
against the set of all feasible requirements. We also compare the
optimal policy (OPT) introduced in \cite{HA01}, which aims to
maximize the total per period reward, $\sum_{X\in
S}q_X/\frac{T}{\tau_X}$. To better illustrate the results, we assume
that all $\hat{q}^*_X$'s are linear functions of two variables,
$\alpha$ and $\beta$. We then find all pairs of $(\alpha, \beta)$ so
that the resulting requirements are fulfilled by the evaluated
policies and plot the boundaries of all such pairs. We call all
pairs of $(\alpha, \beta)$ that are fulfilled by a policy as the
\emph{achievable region} of the policy. We also call the set of all
feasible pairs of $(\alpha, \beta)$ as the \emph{feasible region}.
The complete simulation parameters are shown in Table
\ref{table:simulation:parameter}, in which most parameters are
derived from the simulation set up of \cite{HA01}.

\begin{table*}[tb]
\begin{center}
\begin{tabular}{|c|c|c|c|c|c|c|c|}
\hline
Task id & $\tau_X$ & $m_X$ & $o_X$ & $f^1_X(t)$ & $f^2_X(t)$ & $f^3_X(t)$ & $\hat{q}^*_X$\\
\hline A & 20 & 1 & 10 & $15(1-e^{-t/2})$ & $7\ln(20t+1)$ & $5t$ & $5\alpha$\\
\hline B & 30 & 1 & 15 & $20(1-e^{-3t/2})$ & $10\ln(50t+1)$ & $7t$ & $7\alpha$\\
\hline C & 40 & 2 & 20 & $4(1-e^{-t/2})$ & $2\ln(10t+1)$ & $t$ & $\alpha$\\
\hline D & 60 & 3 & 30 & $10(1-e^{-t/10})$ & $5\ln(25t+1)$ & $4t$ & $4\beta$\\
\hline E & 80 & 4 & 40 & $5(1-e^{-t/2})$ & $3\ln(30t+1)$ & $2t$ & $2\beta$\\
\hline F & 120 & 6 & 60 & $8(1-e^{-t/20})$ & $4\ln(6t+1)$ & $3t$ & $3\beta$\\
\hline
\end{tabular}
\end{center}
\caption{Task parameters for a system in which tasks have different
periods. $f^1_X$, $f^2_X$, and $f^3_X$ correspond to the functions
for exponential, logarithmic, and linear functions, respectively.}
\label{table:simulation:parameter}
\end{table*}

In each simulation of the Greedy Maximizer, we initiate the debt of
$X$ to be $M+1$ and run the simulation for 20 frames to ensure that
it has converged. We then continue to run the simulation for 500
additional frames. The system is considered fulfilled by the Greedy
Maximizer if none of the mandatory parts miss their deadlines in the
500 frames, and the total reward obtained by each task exceeds its
requirements.

\begin{figure*}[t]\subfloat[Exponential functions]{
\label{fig:simulation:exp} 
\includegraphics[width=2.2in]{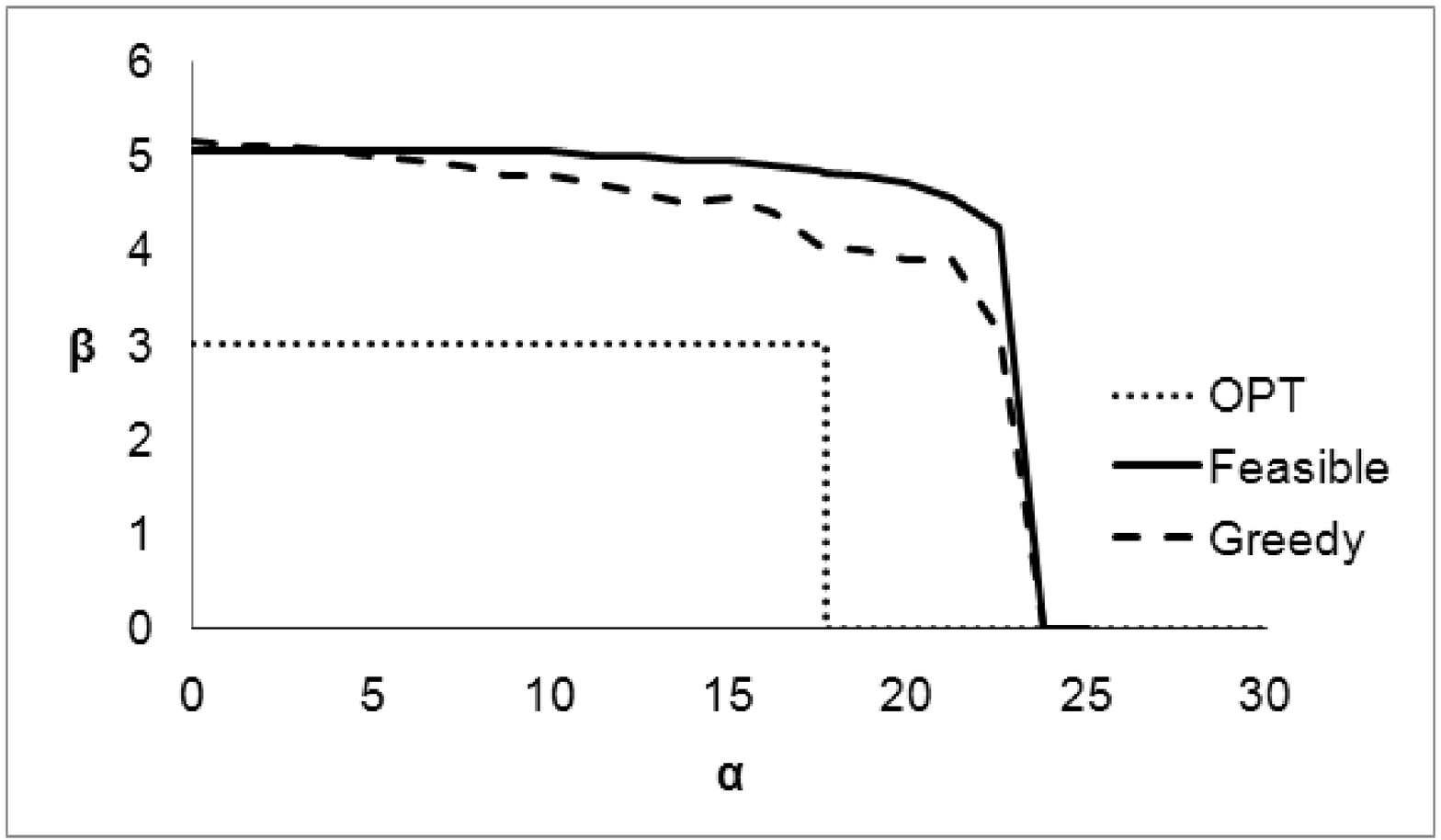}}
\subfloat[Logarithmic functions]{
\label{fig:simulation:log} 
\includegraphics[width=2.2in]{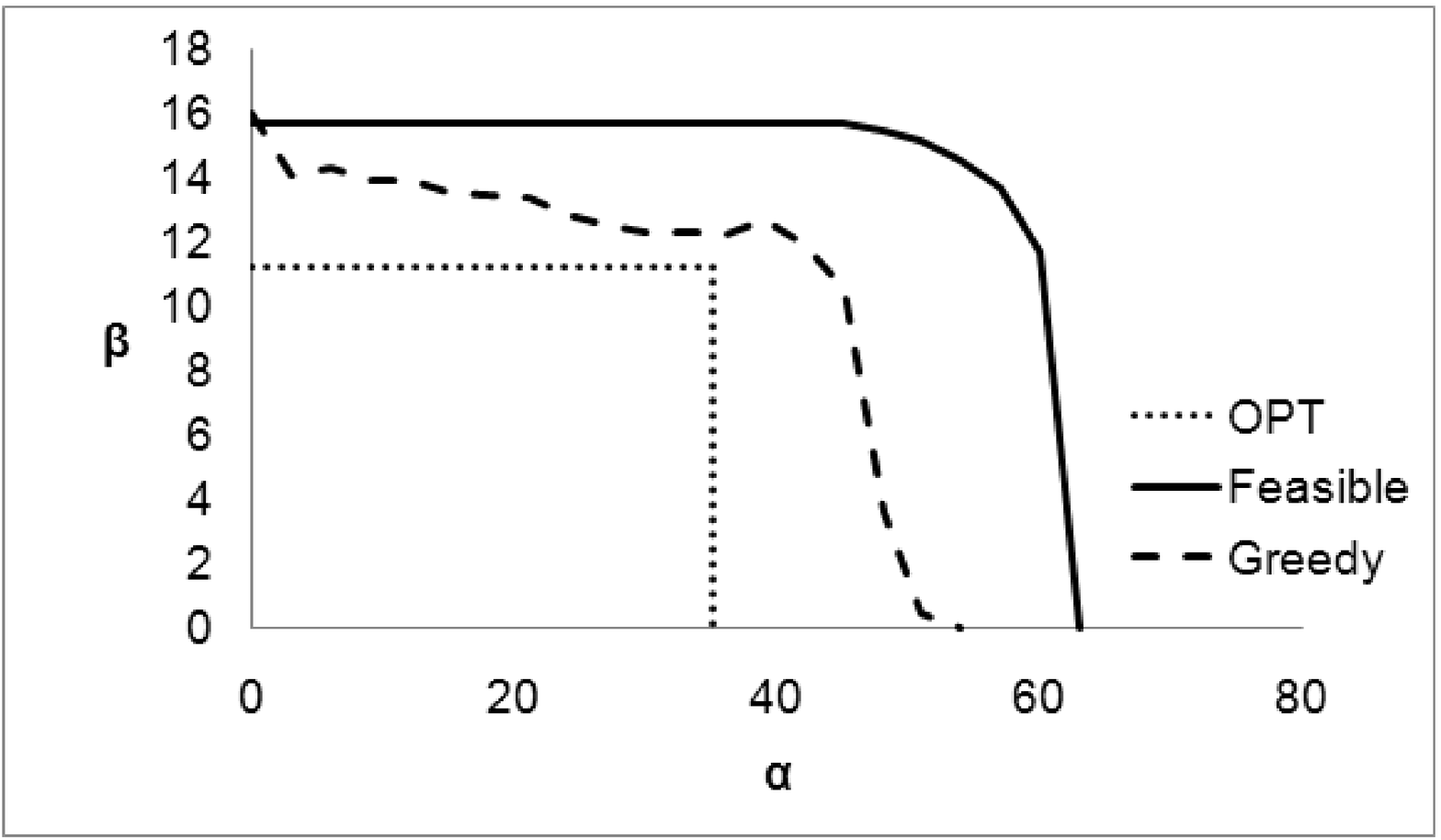}}
\subfloat[Linear functions]{
\label{fig:simulation:linear} 
\includegraphics[width=2.2in]{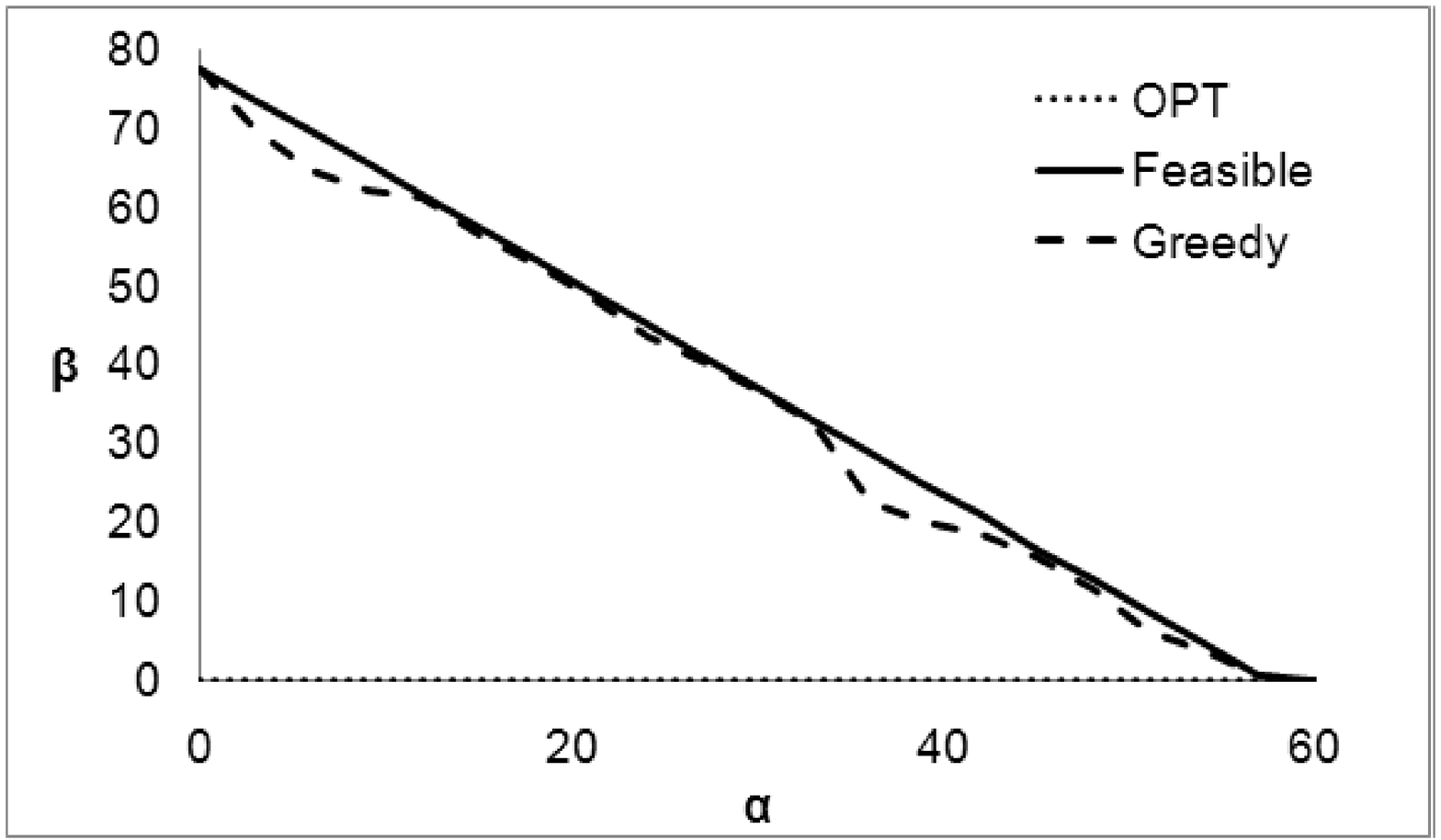}}
\caption{Achievable regions of scheduling policies for the system in
Table \ref{table:simulation:parameter}.}\label{fig:simulation}
\end{figure*}

The simulation results are shown in Fig. \ref{fig:simulation}. For
both cases of exponential and logarithmic functions, the achievable
regions of the OPT policy are rectangles. That is because the OPT
policy only aims at maximizing the total per-period rewards and does
not allow any tradeoff between rewards of different tasks. The
achievable regions of the OPT policy are also much smaller than the
feasible regions. On the other hand, the achievable regions of the
Greedy Maximizer are very close to the feasible region for both the
cases of exponential and logarithmic functions. Also, its achievable
regions are strictly larger than that of the OPT policy. This also
shows that the Greedy Maximizer can provide fine-grained tradeoff
between tasks.

The most surprising result is that for linear functions. In this
case, the OPT policy fails to fulfill any pairs of $(\alpha, \beta)$
except $(0,0)$. A closer examination on the simulation result shows
that, besides mandatory parts, the OPT policy only schedules
optional parts of tasks $D$ and $F$. This example shows that, in
addition to restricted achievable regions, the OPT policy can also
be extremely unfair. Thus, the OPT policy is not desirable when
fairness is concerned. On the other hand, the achievable region of
the Greedy Maximizer is almost the same as the feasible region.
These simulation results also suggest that although we have only
proved that the Greedy Maximizer is a 2-approximation policy, this
approximation bound is indeed very pessimistic. In most cases, the
performance of the Greedy Maximizer is not too far from that of a
feasibility optimal policy.

Next, we simulate a system in which all tasks have the same period.
We assume that $\tau_X=120$, $m_X=0$, and $o_X=120$ for all $X\in
S$. We also simulate all the three functions, exponential,
logarithmic, and linear. Detailed parameters are shown in Table
\ref{table:simulation:parameter2}.

\begin{table}[tb]
\begin{center}
\begin{tabular}{|c|c|c|c|c|}
\hline
Task id & $f^1_X(t)$ & $f^2_X(t)$ & $f^3_X(t)$ & $\hat{q}^*_X$\\
\hline A & $15(1-e^{-t/15})$ & $7\ln(3t+1)$ & $5t$ & $5\alpha$\\
\hline B & $20(1-e^{-3t/8})$ & $10\ln(10t+1)$ & $7t$ & $7\alpha$\\
\hline C & $4(1-e^{-t/5})$ & $2\ln(3t+1)$ & $t$ & $\alpha$\\
\hline D & $10(1-e^{-t/30})$ & $5\ln(15t+1)$ & $4t$ & $4\beta$\\
\hline E & $5(1-e^{-t/5})$ & $3\ln(20t+1)$ & $2t$ & $2\beta$\\
\hline F & $8(1-e^{-t/20})$ & $4\ln(6t+1)$ & $3t$& $3\beta$\\
\hline
\end{tabular}
\end{center}
\caption{Task parameters for a system in which all tasks have the
same period.} \label{table:simulation:parameter2}
\end{table}

\begin{figure*}[t]\subfloat[Exponential functions]{
\label{fig:simulation2:exp} 
\includegraphics[width=2.2in]{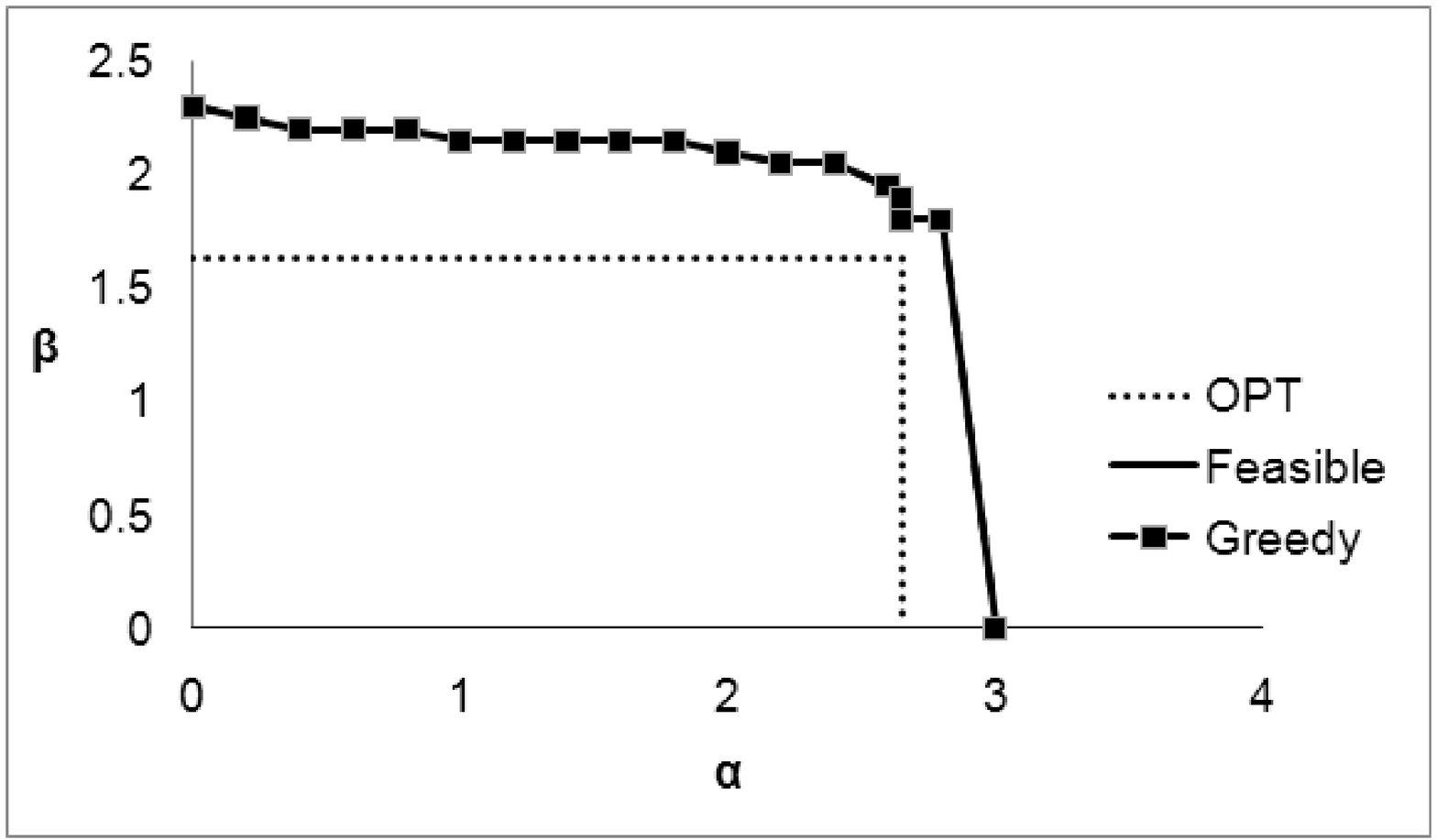}}
\subfloat[Logarithmic functions]{
\label{fig:simulation2:log} 
\includegraphics[width=2.2in]{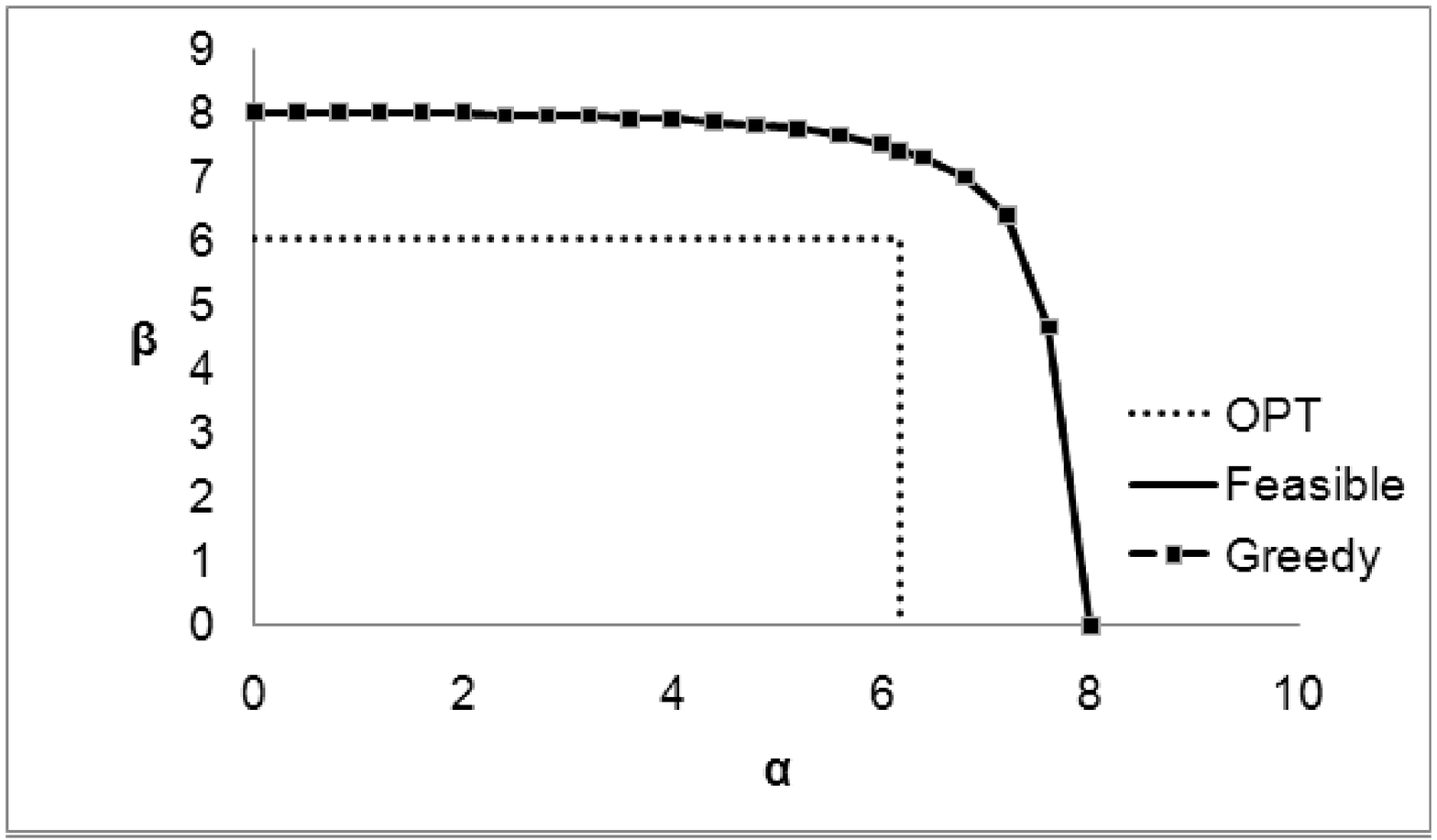}}
\subfloat[Linear functions]{
\label{fig:simulation2:linear} 
\includegraphics[width=2.2in]{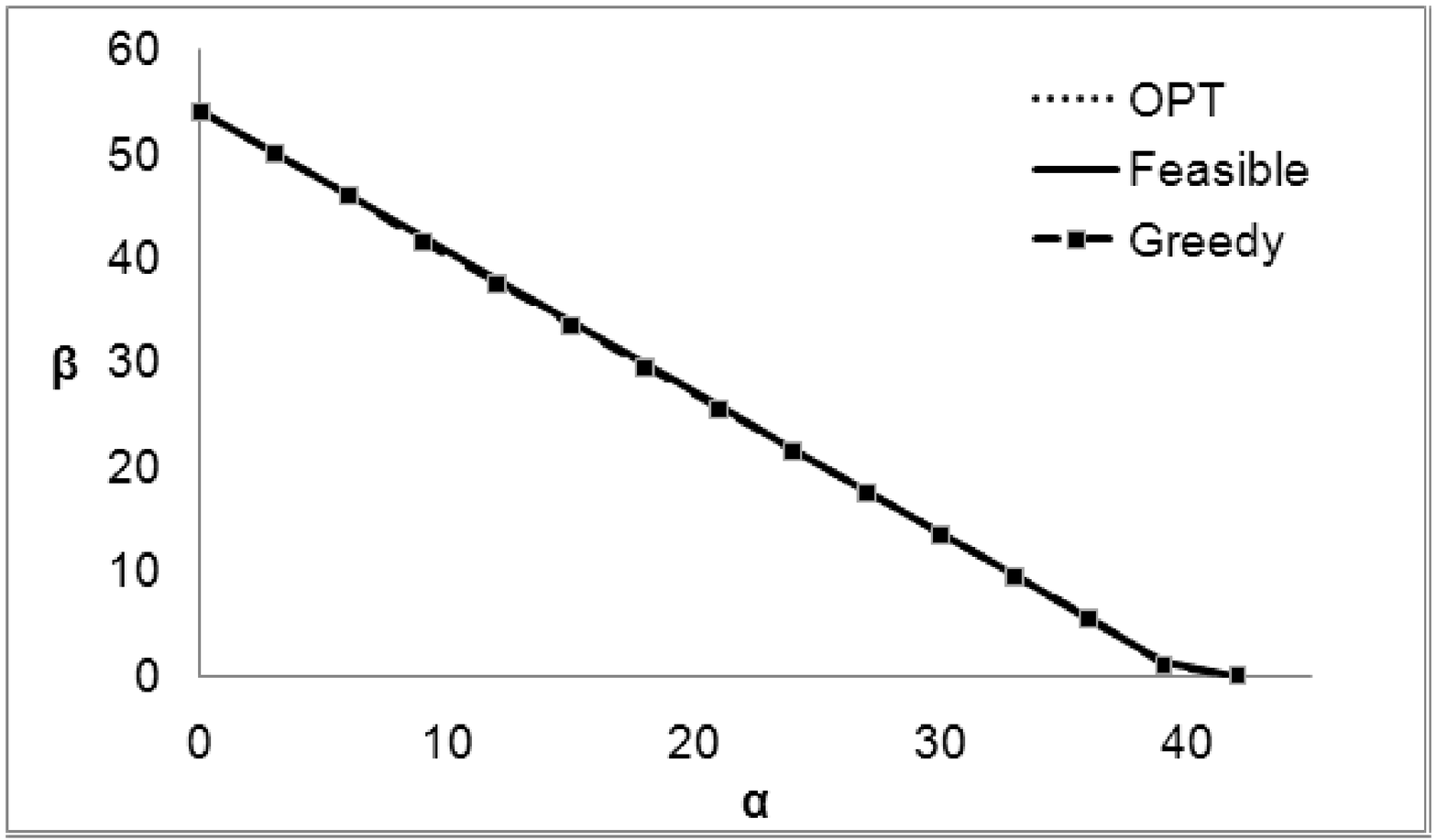}}
\caption{Achievable regions of scheduling policies for the system in
Table \ref{table:simulation:parameter2}.}\label{fig:simulation2}
\end{figure*}

The simulation results are shown in Fig. \ref{fig:simulation2}. As
in the previous simulations, the achievable regions of the Greedy
Maximizer are always larger than those of the OPT policy, for all
functions. Further, the achievable regions of the Greedy Maximizer
are exactly the same as the feasible regions. This demonstrates that
the Greedy Maximizer fulfills every strictly feasible system when
the periods of all tasks are the same.

\section{Concluding Remarks}
\label{section:conclusion}

We have studied a model in which a system consists of several
periodic real-time tasks that have their individual reward
requirements. This model is compatible with both the imprecise
computation models and IRIS models. By making each task specify its
own reward requirement, our model can offer better fairness, and it
allows tradeoff between tasks. Under this model, we have proved a
necessary and sufficient condition for feasibility, and designed a
linear time algorithm for verifying feasibility. We have also
studied the problem of designing on-line scheduling policies and
obtained a sufficient condition for a policy to be feasibility
optimal, or to achieve an approximation bound. We have then proposed
a simple on-line scheduling policy. We have analyzed the performance
of the on-line scheduling policy and proved that it fulfills all
feasible systems in which the periods of all tasks are the same. For
general systems where periods may be different for different tasks,
we have proved that the on-line policy is a 2-approximation policy.
We have also conducted simulations and compared our on-line policy
against a policy that maximizes the total reward in the system.
Simulation results show that the on-line policy has much larger
achievable regions than that of the compared policy.

\section*{Acknowledgement}
The authors are grateful to Prof. Marco Caccamo for introducing us
to this line of work.

\def\baselinestretch{1}
\bibliographystyle{ieeetr}
\bibliography{reference}

\end{document}